\documentclass[11pt]{iopart}
\newcommand{\gguide}{{\it Preparing graphics for IOP journals}}
\usepackage{graphicx}
\usepackage{cite}
\usepackage{parskip}
\usepackage{setspace}
\usepackage{amssymb} 
\usepackage{xcolor}
\usepackage{subfigure}
\usepackage{comment}
\usepackage{ulem}
\usepackage{hyperref}
\usepackage{caption}
\usepackage{comment}

\newcommand{\mycomment}[1]{}
\begin{document}

%\title[Extraction of post-merger GW signals from BNS coalescences]{Detection and characterization of post-merger gravitational wave emission from binary neutron star coalescences}
\title[]{Minimally modeled characterization method of postmerger gravitational wave emission from binary neutron star coalescences}

%\author{M. C. Tringali$^{1,2}$, A. Puecher$^{1,2}$, C. Lazzaro$^3$, R. Ciolfi$^{4,2,5}$, M. Drago$^5$, W. Kastaun$^{5}$, S. Klimenko$^6$, B. Giacomazzo$^{1,2}$, F. Salemi$^5$, G. Vedovato$^3$ ... and G. A. Prodi$^{1,2}$}

\author{M. C.~Tringali$^{1}$, A.~Puecher$^{2,3,*}$, C.~Lazzaro$^{4,5}$, R.~Ciolfi$^{4,6}$, M.~Drago$^{7,8}$, B.~Giacomazzo$^{9,10,11}$, G.~Vedovato$^4$ and G. A.~Prodi$^{12,13}$}

\address{$^1$European Gravitational Observatory (EGO), I-56021 Cascina, Pisa, Italy}
\address{$^2$Nikhef -- National Institute for Subatomic Physics, Science Park 105, 
1098 XG Amsterdam, The Netherlands}
\address{$^3$ Institute for Gravitational and Subatomic Physics (GRASP), Utrecht University, Princetonplein 1, 3584 CC Utrecht, The Netherlands }
\address{$^4$INFN, Sezione di Padova, Via Marzolo 8, I-35131 Padova, Italy}
\address{$^5$Dipartimento di Fisica ed Astronomia, University of Padova, Via Marzolo 8, I-35131 Padova, Italy}
\address{$^6$INAF, Osservatorio Astronomico di Padova, Vicolo dell'Osservatorio 5, I-35122 Padova, Italy}
\address{$^7$Dipartimento di Fisica, University of Rome La Sapienza, Piazzale Aldo Moro 5, I-00185, Italy}
\address{$^8$INFN, Sezione di Roma, Piazzale Aldo Moro 5, I-00185, Italy }
\address{$^9$Dipartimento di Fisica G. Occhialini, Universit\`a di Milano-Bicocca, Piazza della Scienza 3, I-20126 Milano, Italy}
\address{$^{10}$INFN, Sezione di Milano-Bicocca, Piazza della Scienza 3, I-20126 Milano, Italy}
\address{$^{11}$INAF, Osservatorio Astronomico di Brera, Via E. Bianchi 46, I-23807 Merate, Italy}
\address{$^{12}$Dipartimento di Matematica, University of Trento, Via Sommarive 14, I-38123 Trento, Italy}
\address{$^{13}$INFN-TIFPA, Trento Institute for Fundamental Physics and Applications, Via Sommarive 14, I-38123 Trento, Italy}
\vspace{0.2cm}
\address{$^*$Corresponding author: A.Puecher@nikhef.nl}

\begin{abstract}
Gravitational waves emitted during the coalescence of binary neutron star systems carry information about the equation of state describing the extremely dense matter inside neutron stars. In particular, the equation of state determines the fate of the binary after the merger: a prompt collapse to black hole, or the formation of a neutron star remnant that is either stable or survives up to a few seconds before collapsing to a black hole. Determining the evolution of a binary neutron star system will therefore place strong constraints on the equation of state. We present a morphology-independent method, developed in the framework of the coherentWaveBurst analysis of signals from ground-based interferometric detectors of gravitational waves. The method characterizes the time-frequency postmerger gravitational-wave emission from a binary neutron star system, and determines whether, after the merger, it formed a remnant neutron star or promptly collapsed to a black hole. We measure the following quantities to characterize the postmerger emission: ratio of signal energies and match of luminosity profile in different frequency bands, weighted central frequency and bandwidth. From these quantities, based on the study of signals simulated through injections of numerical relativity waveforms, we build a statistics to discriminate between the different scenarios after the merger. Finally, we test our method on a set of signals simulated with new models, to estimate its efficiency as a function of the source distance.
%finding an efficiency of \anna{compute} over the events for which the postmerger signal is detected.

\end{abstract}

\mycomment{
\textcolor{red}{add here possible titles:\\
\noindent TITLE 1 - Characterization of Neutron Star remnants by gravitational waves from the post-merger phase after coalescence\\
TITLE 2 - Detection and characterization of post-merger gravitational wave emission from binary neutron star coalescences\\
TITLE 3 - A novel approach to the extraction of post-merger gravitational wave signals in binary neutron star coalescences\\
TITLE 4 - A new analysis tool to extract post-merger gravitational wave signals in binary neutron star coalescences\\
TITLE 5 - Extracting post-merger gravitational wave signals in binary neutron star coalescences\\
}
}

% Comment out if separate title page not required\maketitle

\section{Introduction}
% INTRODUCTION
%\textcolor{red}{Ric: drafting the intro (post-BNS point of view), work in progress..}
%The observations results of Advanced LIGO and Advenaced Virgo during O1 and O2 data taking have been reported in the GWTC-1 catalog \cite{GWTC-1}. The first detection of GW signal due to a binary neutron star (BNS) coalescence \cite{GW170817disc}, GW170817, 

\renewenvironment{comment}{}{}
%\newcommand{\mycomment}[1]{}
%basic statements on GWastronomy ref detectors cataloghi and Multimesssenger observation

\noindent

During the third observing run (O3), Advanced LIGO \cite{Advanced_LIGO, LIGOScientific:2014pky} and Advanced Virgo \cite{Advanced_Virgo, VIRGO:2014yos} interferometers detected 79 events, bringing the number of gravitational wave (GW) events observed to a total of 90. All the GW signals detected up to now were generated by the coalescence of compact objects \cite{GWTC-1, Abbott_2020niy, GWTC-3}, and among them two have been classified as coalescences of binary neutron star (BNS) systems: GW170817 \cite{TheLIGOScientific:2017qsa} and GW190425 \cite{Abbott:2020uma}. GW170817 marked the beginning of the so-called multimessenger astronomy era, since, for the first time, we observed the GWs emitted by the merger of two neutron stars (NSs) and the electromagnetic counterpart signal in different energy bands \cite{LIGOScientific:2017ync}.
BNS mergers offer us a tool to study the equation of state (EoS) of supranuclear dense matter, through the imprint that it leaves on the GW signal emitted during coalescence. GWs emitted during the inspiral phase allow placing constraints on the EoS through measurements of the tidal deformability parameter \cite{Dietrich:2020eud,Hinderer:2009ca, Damour:2012yf, DelPozzo:2013ala,Lackey:2014fwa, 
Agathos:2015uaa, Dietrich:2018uni}, as was done for GW170817 \cite{LIGOScientific:2018hze}. 
Moreover, the EoS determines the fate of the BNS among the possible scenarios for the system evolution: prompt collapse to black hole (BH), formation of a hypermassive (HMNS) or supramassive (SMNS) neutron star with a delayed collapse to black hole, or even a stable NS. Each of these scenarios affects differently the GW signal emitted during the postmerger phase \cite{Piro:2017zec, Sarin:2020gxb}. Therefore, the characterization of the system evolution and of the spectral properties of the postmerger GW signal yields information about the EoS that is complementary to what can be inferred from the inspiral GW emission.
For GW170817, a targeted search for postmerger GW emission was performed, looking for both short ($\lesssim 1 s $) \cite{LIGOScientific:2017fdd} and intermediate (up to 500 s) \cite{LIGOScientific:2018urg} duration signals. No evidence of a postmerger emission was found, but results showed that with the sensitivity of Advanced LIGO and Advanced Virgo the source distance should have been at least one order of magnitude less for the postmerger signal to be detectable.
Current detectors are strongly limited by quantum shot noise in the kilohertz region, which is where the postmerger GW emission is expected to lie. However, postmerger detections are expected to become feasible with the future improvements planned for Advanced Virgo+ and Advanced LIGO+ \cite{Virgosqueezing, LIGOsqueezing}, and especially with third-generation detectors, like Einstein Telescope \cite{Punturo:2010zz, Maggiore:2019uih, Freise:2008dk, Hild:2009ns,Sathyaprakash:2011bh} and Cosmic Explorer \cite{Reitze:2019iox, Evans:2021gyd}. 
Due to the variety and complexity of the physical processes involved, which are not fully understood yet, it is hard to get exact theoretical models of the postmerger signal morphologies, which cover a wide range of possible spectral features. Therefore, model-independent analyses, like the ones proposed in \cite{Clark:2014wua, Clark:2015zxa, Chatziioannou:2017ixj}, are particularly suited to infer properties of the postmerger GW emission. 
Efforts have also been made to construct postmerger models, usually based on the study of numerical relativity (NR) simulations and on quasi-universal relations \cite{Bauswein:2011tp, Takami:2014zpa, Rezzolla:2016nxn, Bernuzzi:2015rla, Bauswein:2012ya, Hotokezaka:2013iia, Bauswein:2014qla, Takami:2014tva, Bauswein:2015yca, Lioutas:2021jbl}, both in time \cite{Breschi:2019srl, Easter:2020ifj, Soultanis:2021oia} and frequency domain \cite{Breschi:2022xnc,Puecher:2022oiz}. In \cite{Wijngaarden:2022sah}, the authors present a hybrid model to describe the GW signal emitted during the coalescence of a BNS system, employing analytical templates to describe the pre-merger phase, and a morphology-independent approach, based on sine-gaussian wavelets, to analyze the postmerger signal. With a different approach, \cite{Easter:2018pqy} presents a hierarchical model to generate postmerger spectra based on numerical relativity simulations.
In \cite{Clark:2014wua}, the authors introduce a morphology-independent method to search for postmerger signals, and a model-selection and parameter-estimation algorithm to distinguish, based on the reconstructed waveform, between the possible postmerger scenarios.

In this work, we develop, in the framework of the coherentWaveBurst (cWB) algorithm \cite{Klimenko:2015ypf, Klimenko:2008fu, Klimenko:2005xv, klimenko_sergey_2021_4419902, Drago:2020kic}, a model-independent analysis tool to characterize the postmerger signal emitted by a BNS system, and to use the signal's features to determine whether the BNS merger results in a prompt collapse to BH or creates a metastable massive NS, in which case a postmerger signal is present. Our method does not employ a specific waveform model, nor it depends on the assumption of a single gaussian-shaped dominant peak in the postmerger spectrum, as was done for example in \cite{Clark:2014wua}.

The paper is organized as follows: Sec.~\ref{sec:nr_waveforms} describes the catalog of NR waveforms employed in this work and the simulation settings, Sec.~\ref{sec:params} explains the quantities and method used to characterize the postmerger signal, and in Sec.~\ref{sec:statistics} we illustrate the procedure to discriminate between the different postmerger scenarios. Results are shown in Sec.~\ref{sec:results}, and conclusions are reported in Sec.~\ref{sec:conclusions}.

\section{Simulated signals}
\label{sec:nr_waveforms}
% CATALOG

\begin{table}[tb]
\footnotesize

\centering
%\begin{center}
\begin{tabular}{clcccccc}
\hline
& Model & Ref. & $M_\mathrm{b}$ [$M_{\odot}$] & $M_{\infty}$ [$M_{\odot}$] & $\tau_\mathrm{MNS}$ [ms] & $M_\mathrm{BH}$ [$M_{\odot}$] & f$_\mathrm{peak}$ [kHz] \\
\hline
%&\color{red}{SHT-M2.0-I} & \cite{Kastaun15} & 4.01 & 1.80 & $>16.9$ & ... & 2.47 \\
&SHT-M2.0-S & \cite{Kastaun15} & 4.01 & 1.80 & $>9.4$ & ... & 2.66 \\
&SHT-M2.2-I$^\bullet$ & \cite{Kastaun15} & 4.39 & 1.95 & $<1$ & 3.73 & ... \\
%&\color{red}{LS220-M1.5-I} & \cite{Kastaun15} & 3.12 & 1.41 & 8.6 & 2.65 & 3.24 \\
&LS220-M1.5-S & \cite{Kastaun15} & 3.12 & 1.41 & 7.7 & 2.67 & 3.17 \\
&LS220-M1.7-I$^\bullet$ & \cite{Kastaun15} & 3.46 & 1.54 & $<1$ & 2.98 & ... \\
%&\color{red}{LS220-M1.8-I} & \cite{Kastaun15} & 3.62 & 1.61 & $<1$ & 3.14 & ... \\
&APR4-HM$^\bullet$ & \cite{Endrizzi16} & 3.18 & 1.43 & 1 & 2.79 & ... \\
&{H4-q10-Kaw} & \cite{Kawamura16} & 3.04 & 1.40 & 12 & 2.47 & 2.67 \\
%&\color{red}{H4-q08-Kaw} & \cite{Kawamura16} & 3.04 & 1.54,1.26 & 25 & 2.50 & 2.69 \\
%&\color{red}{H4-q10} & \cite{Ciolfi17} & 2.92 & 1.35 & 22 & 2.49 & 2.54 \\
%&\color{red}{H4-q09} & \cite{Ciolfi17} & 2.92 & 1.42,1.29 & 28 & 2.42 & 2.55 \\
&BL & \cite{Endrizzi18} & 2.95 & 1.35 & $>20$ & ... &  3.17 \\
&\textbf{LS220-M1.5-I}& \cite{Kastaun15} & 3.12 & 1.41 & 8.6 & 2.65 & 3.24 \\
&\textbf{H4-q08-Kaw} & \cite{Kawamura16} & 3.04 & 1.54,1.26 & 25 & 2.50 & 2.69 \\
\hline
%&\color{red}{APR4-LM} & \cite{Endrizzi16} & 2.66 & 1.22 & $\mathrm{SMNS}$ & ... & 3.17 \\
&APR4-UM  & \cite{Endrizzi16} & 3.01 & 1.42,1.29 & $\mathrm{SMNS}$ & ... & 3.30 \\
%&\color{red}{SHT-M1.5-I} & \cite{Kastaun16} & 3.03 & 1.40 & $\mathrm{stable}$ & ... & 2.04 \\
%&\color{red}{APR4-q10} & \cite{Ciolfi17} & 2.98 & 1.35 & $\mathrm{SMNS}$ & ... & 3.35 \\
&APR4-q09 & \cite{Ciolfi17} & 2.98 & 1.42,1.28 & $\mathrm{SMNS}$ & ... & 3.24 \\
&APR4-1.35-Long & \cite{Ciolfi19} & 2.98 & 1.35 & $\mathrm{SMNS}$ & ... & 3.35 \\
%&\color{red}{MS1-q10} & \cite{Ciolfi17} & 2.91 & 1.35 & $\mathrm{stable}$ & ... & 2.03 \\
%&\color{red}{MS1-q09} & \cite{Ciolfi17} & 2.91 & 1.41,1.28 & $\mathrm{stable}$ & ... & 2.09 \\
&\textbf{APR4-LM} & \cite{Endrizzi16} & 2.66 & 1.22 & $\mathrm{SMNS}$ & ... & 3.17 \\
\hline
%\hline
%\hline
%&LS220-M1.5-I & \cite{Kastaun15} & 3.12 & 1.41 & 8.6 & 2.65 & 3.24 \\
%&H4-q08-Kaw & \cite{Kawamura16} & 3.04 & 1.54,1.26 & 25 & 2.50 & 2.69 \\
%&APR4-LM & \cite{Endrizzi16} & 2.66 & 1.22 & $\mathrm{SMNS}$ & ... & 3.17 \\
%\hline
\end{tabular}
\caption{\small{Parameters of BNS models: total baryonic mass ($M_\mathrm{b}$), gravitational mass of each NS at infinite separation ($M_{\infty}$), remnant lifetime and final BH mass ($\tau_\mathrm{MNS}$, $M_\mathrm{BH}$; only for HMNS remnants, upper part of the Table), frequency of the dominant peak in the post-merger GW signal ($f_\mathrm{peak}$). For cases with prompt collapse to a BH (remnant lifetime $\lesssim 1$~ms) no estimate of $f_\mathrm{peak}$ is available. %The first two BNS models listed form a HMNS remnant, but the simulations end before the collapse to a BH. 
The BNS models in the upper part of table for which no $\tau_{MNS}$ and $M_{BH}$ are available form a HMNS remnant, but the simulations end before the collapse to BH.
Model labels are chosen in order to facilitate the correspondence with the original references. 
The two models with suffix ``-S" have initial spin, while all the others are irrotational. %In this waveforms sample, we did not include models that form a stable NS after the merger, since they are not expected to emit a post-merger gravitational-wave signal, therefore are not relevant for our analysis.
The models with names in bold are the ones used to test the performance of the method, the other ones are instead used to tune the procedure. The models with a $^\bullet$ after their name are the ones with prompt collapse to BH.}}
\label{tab:NRcatalog}
%\end{center} 
\end{table}

For our analysis, we employ simulated signals produced by injecting NR waveforms in Gaussian noise, assuming an Advanced LIGO-Advanced Virgo network with the design sensitivity planned for O4 \cite{KAGRA:2013rdx} \footnote{We note that the O4 design sensitivity curves have been updated \cite{KAGRA:2013rdx} with respect to the ones that were available when we started this work. However, the sensitivity curves are equivalent in the frequency range of interest for this work.}.
We consider a sample of 13 waveforms obtained from NR simulations of BNS mergers {\cite{Kastaun15,Endrizzi16,Kawamura16,Kastaun16,Ciolfi17,Endrizzi18,Ciolfi19}};
among those, ten are used to tune the method and three to test the performance of our analysis pipeline. 
If the system total baryonic mass is smaller than the maximum mass supported by a uniformly rotating NS, the merger remnant is either indefinitely stable, or  a SMNS that collapses to BH on long timescales, $\gg1$~s; if, instead, the total mass is larger, the system collapses directly to a BH, or forms a short-lived HMNS, which collapses to BH within a few tens of ms after the merger. 
Among our sample, three models are characterized by a prompt collapse to BH, and thus they include no postmerger GW signal other than the very weak and short one produced by the ringdown of the resulting BH. For all the other waveforms, a SMNS or HMNS is formed, and hence a postmerger signal is present, even if with different durations and morphologies. We did not include models that form a stable NS after the merger.
%, since they are not expected to emit a significant GW transient in the postmerger, and therefore are not relevant for this study.
The main features of the models in our waveform sample are provided in Table~\ref{tab:NRcatalog}. %~\ref{NRcatalog}. 
They include different masses and mass ratios, spanning a wide range of NS compactness. Five different EoS are considered: APR4 \cite{APR4}, %MS1 \cite{MS1}, 
H4 \cite{H4}, LS220 \cite{LS220,Kastaun15}, SHT \cite{SHT} and BL \cite{BLeos}. Some of the models from \cite{Kastaun15} have also initial spin %(i.e.,~they are not irrotational), 
while the models from \cite{Kawamura16,Ciolfi17} are magnetized, although the effects of the corresponding magnetic field are negligible for the purposes of the present analysis. 
Details and specific calculations for each model can be found in the references indicated in Table~\ref{tab:NRcatalog}.

Injections are performed with the cWB simulation engine tool \cite{cwb_inj}, using, for simplicity, only the dominant oscillation mode $(\ell ,m)=(2,2)$ GW signal extracted from the NR waveforms. Following the Newman-Penrose formalism \cite{Baiotti2008}, the corresponding plus and cross polarization GW signals are given by
\begin{equation}
h_{+,\times} = \sum_{\ell =2}^{\infty}  \sum_{m =-\ell}^{\ell} h_{+,\times}^{\ell m} ~ _{-2}Y_{\ell m}(\theta, \phi)\approx h_{+,\times}^{22}~_{-2}Y_{22},
\end{equation}

\noindent
where $ _{s}Y_{lm}(\theta, \phi)$ are the spin-weighted spherical harmonics \cite{Goldberg}.

\begin{figure}[ht!]
\begin{center}
{\includegraphics[width=0.45\textwidth]{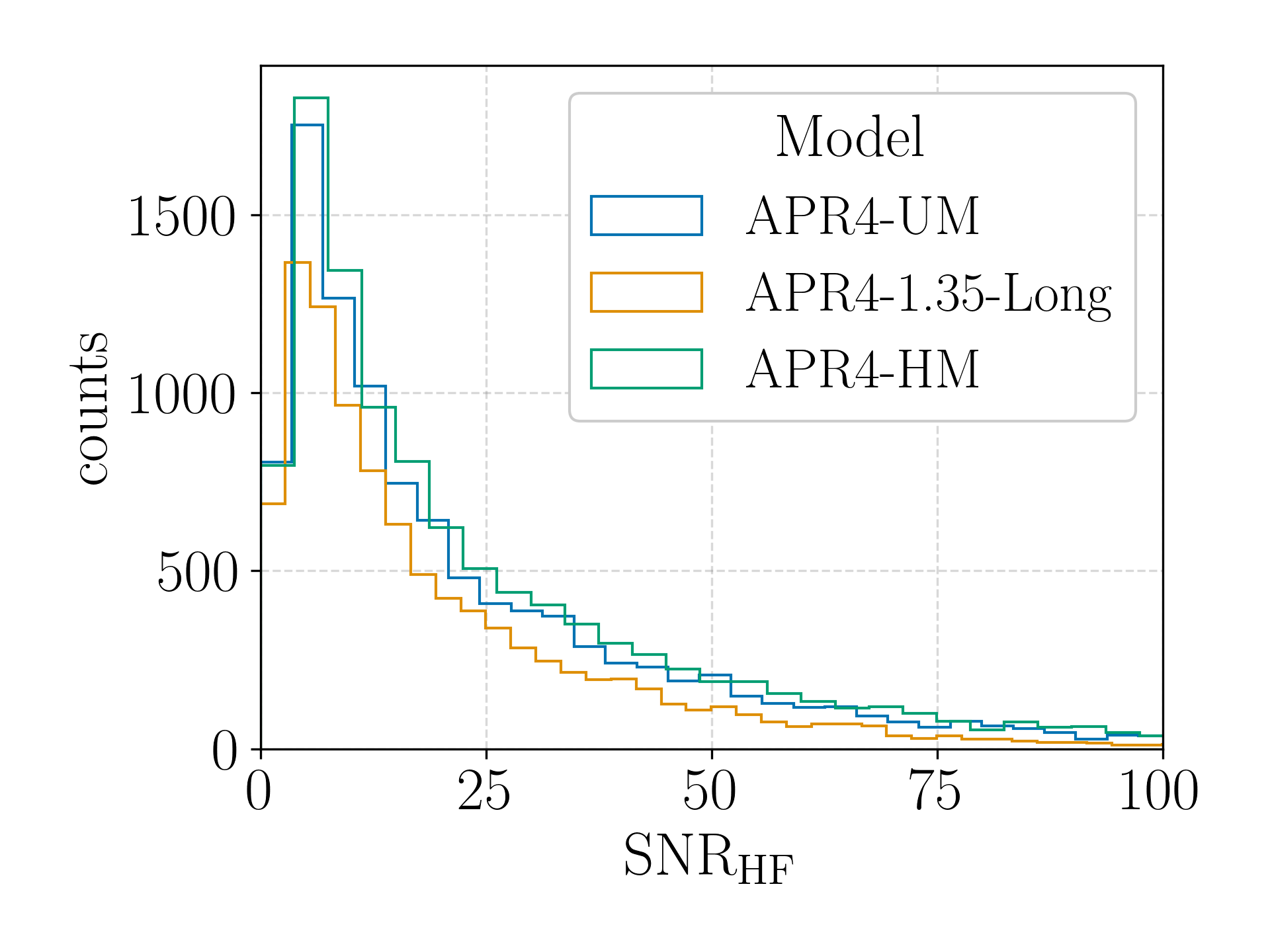}} 
{\includegraphics[width=0.45\textwidth]{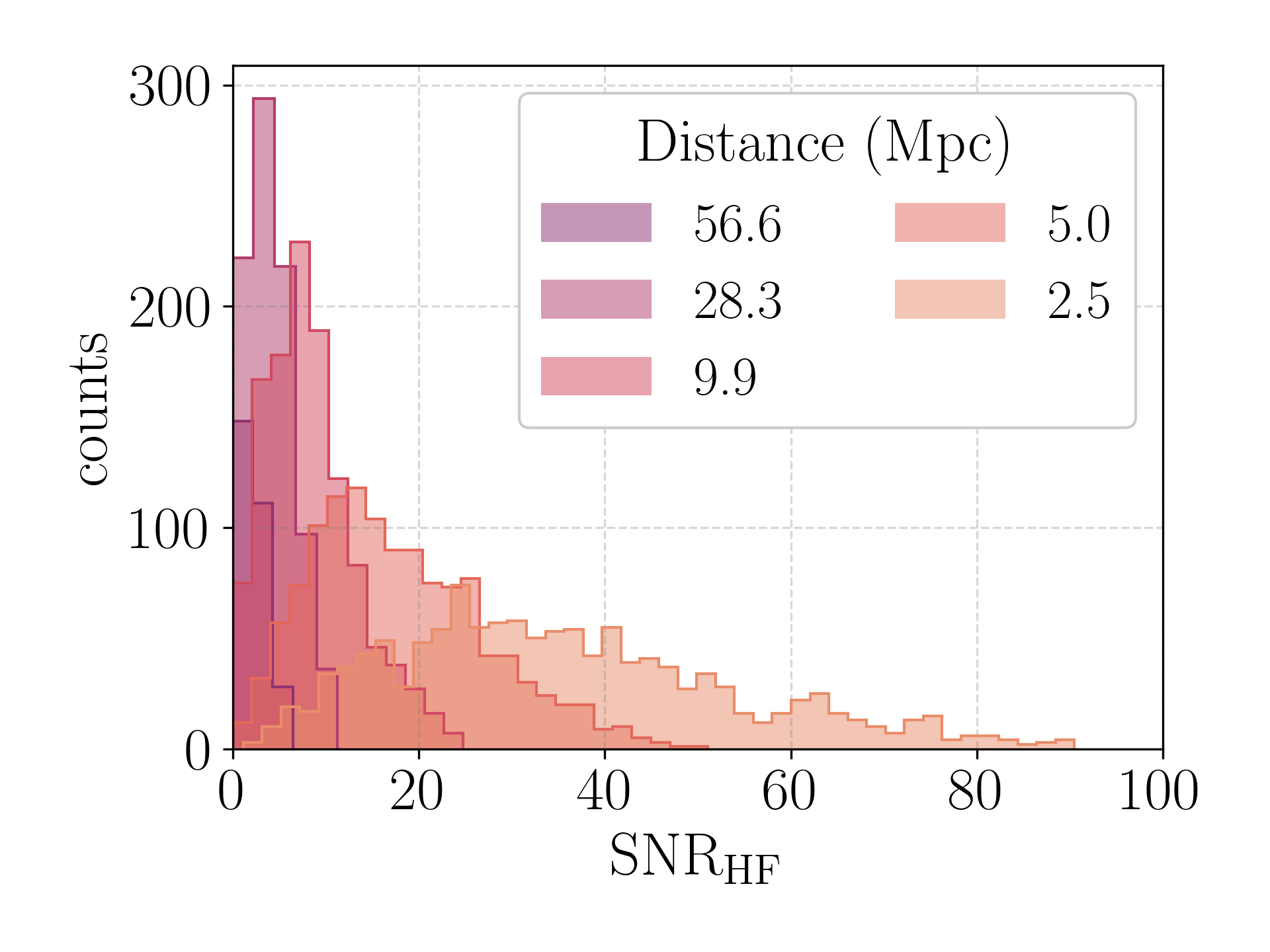}}
\caption{\small{Left panel: distribution of $\rm SNR_{HF}$ for different models considered, for all simulated events. Right panel: distribution of $\rm SNR_{HF}$ for the SHT-M2.0-S model for injections at different distances.}}

\label{fig:snr_hf_dis_pmns}
\end{center}
\end{figure}

We consider a source population on a grid of fixed distances between $[2.50,71.18]$~Mpc. For each distance, injections are distributed isotropically in the sky and with a random distribution of the possible BNS system orientation. In total, we simulate roughly $\rm 10k-15 k$ events for each model. \\
Since we are mainly interested in the merger and postmerger part of the signal, the injected waveforms include only a few late cycles of inspiral.
Therefore, the signal-to-noise ratio (SNR) of our simulated signals does not correspond to what would be measured by a real detector with the whole inspiral contribution. For this reason, we define a new parameter to characterize the reconstructed signal energy, the high-frequency signal-to-noise ratio $\rm SNR_{HF}$, i.e., the cWB reconstructed SNR for frequencies higher than 768 Hz. The reason for the choice of this specific frequency value is related to the analysis settings, as explained in Sec.~\ref{sec:pm-analysis}.
Figure \ref{fig:snr_hf_dis_pmns} shows the distribution of $\rm SNR_{HF}$ over all the simulated events used to tune our procedure, for the different waveform models (left panel), and for one specific model but for different source distances (right panel).

 %~\ref{NRcatalog} 

%\section{Reconstruction of information carried by GW}
\section{Reconstruction and characterization of the postmerger signal}
\label{sec:params}
% DA METHODS

\subsection{cWB pipeline overview}

Searches based on general methodologies, which in particular do not assume any specific time evolution for the signal's phase and amplitude, can be really profitable to extract physical knowledge for postmerger events.
An example of such morphology-independent search pipeline is the cWB algorithm \cite{Klimenko:2015ypf, Klimenko:2008fu, Klimenko:2005xv, klimenko_sergey_2021_4419902, Drago:2020kic},
which is based on a constrained likelihood maximization of the detector network response, performed in time-frequency (TF) domain by using the Wilson-Daubechies-Meyer wavelet transforms. The cWB algorithm selects the TF pixels with a significant energy excess and coherent power across the detectors, combining multiple TF resolutions.
%cWB repeats the selection of time-frequency pixels with significant energy excess and coherent power across the detectors at several time-frequency resolutions, so to build a representation of GW candidate events by a cluster of the most significant multi-resolution pixels. 
The likelihood statistic is built as a coherent sum over the pixels selected, and it is maximized on a sky loop of the possible positions of the source; finally the maximum of the likelihood statistic is used to reconstruct the signal properties. %, such as the waveform, the source sky location etc.

In this work, we develop a follow-up analysis, i.e., we assume that the cWB pipeline already reconstructed the main event and we search for possible postmerger signals. To enhance the reconstruction capability of the cWB algorithm on the postmerger phase of the signal, we set up an optimized configuration of the pipeline: the follow-up postmerger search is performed in the frequency band 512-4096~ Hz. %, employing frequency resolutions in the range $[4,256]$~Hz, in steps of $\rm 2^n$, which correspond to time resolutions in the range $[1.9,125]$ ~ms. 
The pipeline has been optimized extending the separation, both in time and frequency, allowed between pixels included in the same cluster, to favor the grouping of the postmerger phase pixels in the trigger.

\subsection{Follow-up analysis of the postmerger phase}
\label{sec:pm-analysis}

The procedure developed inspects the morphology of postmerger signals in TF domain, with the aim of discriminating the two possible scenarios after the BNS coalescence, namely a {\it prompt} or {\it delayed collapse} to BH; in the latter one, the formation of a massive NS remnant is predicted, which emits a GW signal in high frequency band, not expected in the case of prompt collapse.

The analysis is performed on the TF map (spectrogram) of the reconstructed signal computed separately for each detector. In particular, all the results shown in the following refer to the coherent GW candidate reconstructed by cWB in the Hanford-Livingston-Virgo network, and then projected on the Livingston detector. The first step consists in identifying on each TF map the region in which we expect to find the potential postmerger part of the signal. This is accomplished by defining two cuts, one in frequency and one in time, as introduced in \cite{tesi_Tringali}, on TF maps with a frequency and time resolution $\rm df = 512~Hz$ 
%\footnote{Each bin in frequency is called \textit{layer}, and due to technicalities of the cWB algorithm the first layer corresponds to only half bin, i.e., for this resolution it spans the frequency range [512,768] Hz.} 
and $\rm dt = 0.98~ms$, to provide the needed resolution in time domain.

The frequency cut is chosen a priori at $\rm f_{cut} = 1792~Hz$\footnote{In the Wilson-Daubechies-Meyer transform used by cWB, the first bin is set to have a bandwidth corresponding to half the frequency resolution, thus, in our case, it spans the range [0, 256 Hz], and the next bin boundaries are at 768, 1280, 1792 Hz etc.}, considering this choice the most appropriate bin-edge value to separate the upper frequency region in which the postmerger signal can be expected \cite{Rezzolla2016, Bauswein2015} from the lower frequency region of the late inspiral and merger emission. The time cut $\rm t_{cut}$, instead, is set as the energy-weighted mean time in the frequency band $[768, 1280]$~Hz, which provides an estimate of the time at which the merger occurred.

\begin{figure}[ht!]
\begin{center}

\includegraphics [width=0.7\textwidth]{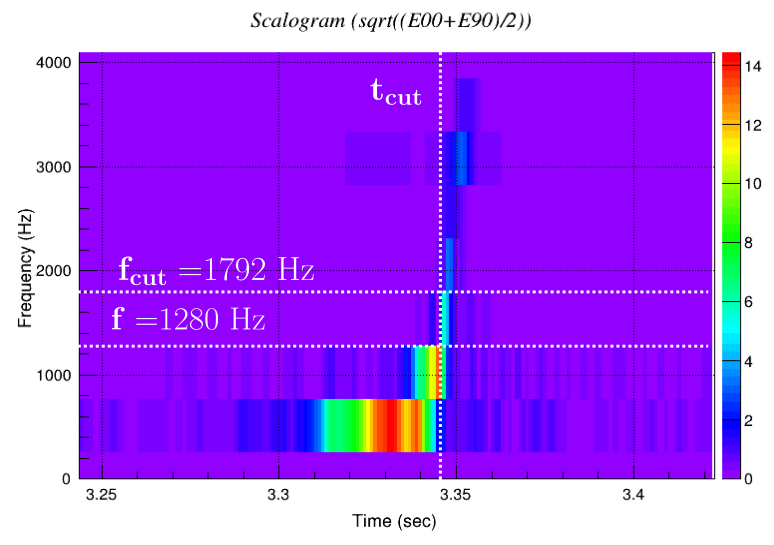}
\end{center}
\caption[]{\small{Example of TF map for the reconstructed GW signal including postmerger of an event simulated with the SHT-M2.0-S model. The TF map shows the division in four quadrants: the vertical and horizontal dashed lines represent $\rm f_{cut}$ and $\rm t_{cut}$, set as explained in the text. We show also the $\rm f=1280~Hz$ boundary that is used to determine the late-merger region and $\rm t_{late}$ for the pre-selection criteria.}}
\label{tfmap_quad}
\end{figure}

 The TF map results thus divided as reported in Fig.~\ref{tfmap_quad}: we identify as postmerger region (PM) the one marked by $\rm t > t_{cut}$ and $\rm f>f_{cut}$; the late merger (LM) emission is expected instead to be in the region with $ \rm t>t_{cut}$ and frequency in range [1280, 1792]~Hz.

Since the procedure aims to characterize the postmerger signal, we want to apply it only to events for which a GW signal after the merger might be detected. We translate this into two conditions: 

\noindent\begin{minipage}{0.45\textwidth}
\begin{equation*}
\rm SNR_{rec}>10
%\label{eq:precut1}
\end{equation*}	
\end{minipage}%
\begin{minipage}{0.45\textwidth}
\begin{equation}
\rm t_{late} > \rm t_{cut},
\label{eq:precut}
\end{equation}
\end{minipage}

where $\rm SNR_{rec}$ is the network reconstructed SNR above 768~Hz, $\rm t_{late}$ is the energy-weighted time for $\rm f>1280$~Hz and $\rm t > t_{cut}$. % the energy-weighted time for $\rm 768 < f < 1280 $~Hz.

\begin{figure}[ht!]
\begin{center}
{\includegraphics[width=0.9\textwidth]{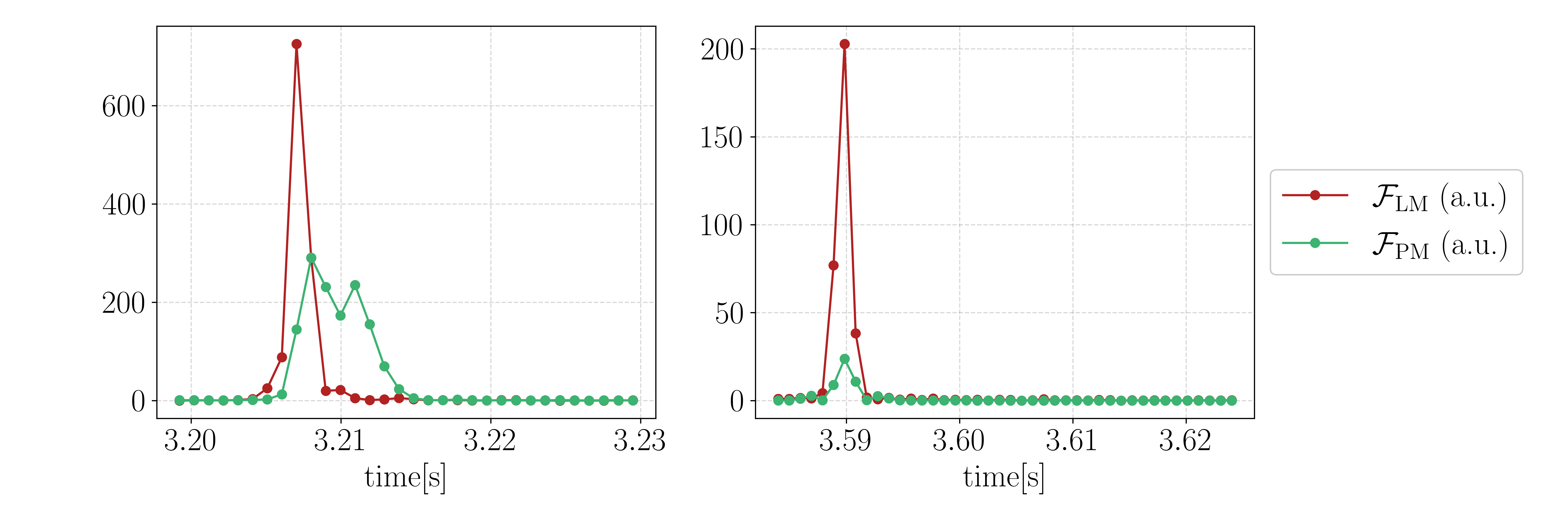}} 
\caption{\small{Luminosity profile for two sample events, on the left for the SHT-M2.0-S model, expected to form a massive NS remnant, on the right for SHT-M2.2-I, expected to promptly collapse to a BH. $\mathcal{F}_{\rm LM}$ is shown in red, $\mathcal{F}_{\rm PM}$ in green.}} 
\label{fig:lum_profile}
\end{center}
\end{figure}

For the events satisfying both these pre-selection criteria, we estimate the following physical quantities that describe time and frequency features of the postmerger signal. All the listed parameters are estimated from the reconstructed GW signal without any a-priori assumption on the source or on the signal morphology:

\begin{itemize}
 
\item ratio between energy in the postmerger region ($\rm E_{PM}$) and energy in the late-merger region ($\rm E_{LM}$)

\begin{equation}
   \rm E_{ratio} = \log_{10} \left( \frac{E_{PM}}{E_{LM}}\right)
\end{equation}

\item luminosity profile overlap function $\mathcal{O}$, defined as:

\begin{equation}
\mathcal{O}=  \frac{ \sum_{i} \mathcal{F}_{\rm LM}[i]  \mathcal{F}_{\rm PM}[i]} {\left( \sum_{i}\mathcal{F}_{\rm LM}[i]^2 \sum_{i}\mathcal{F}_{\rm PM}[i]^2 \right)^{1/2}},
\end{equation} 

where the luminosity profile function $\mathcal{F}[i]$ is
%is estimated as the amount of 
the energy emitted per fixed interval of time, which is defined by the TF map time resolution, and the sum $\sum_{i}$ is performed over the times of pixels after $\rm t_{cut}$. The luminosity profile
$\mathcal{F}_{\rm LM}$ is calculated in the frequency range  1280-1792~Hz, while $\mathcal{F}_{\rm PM}$ in the frequency range  1792-4096~Hz. Figure~\ref{fig:lum_profile} shows an example of luminosity profile function for an event expected to form a NS remnant, on the left, and one with prompt collapse to BH, on the right. In the case of formation of a remnant, we can clearly identify two peaks in the luminosity profile function, one due to the luminosity of the GW signal in the late-merger phase, and one due to the luminosity of the PM signal. On the other hand, the PM peak is almost not present in the plot on the right, consistently with what we expect in the case of a prompt collapse to BH. \\

%The total luminosity profile function shows two clearly distinct peaks in the case of formation of a remnant, one due to the luminosity of the GW signal in the late-merger phase, and one due to the luminosity of the PM signal. On the other hand, the PM peak is almost not present in the plot on the right, consistently with what we expect in the case of a prompt collapse to BH. \\
\noindent

\end{itemize}

For the computation of spectral features, we consider TF maps of reconstructed events with a better frequency resolution $\rm df = 64~Hz$, which implies $\rm dt \simeq 7.8~ms$. 
%Regarding the cuts for the PM region identification on TF maps $t_{cut}$ is kept at the one previously found withthe finest time step, while the frequency cut is now taken at $f'_{cut}= 1760 Hz$  in order to keep it a layer edge
The PM region is here defined by the same $\rm t_{cut}$ as before, but the frequency threshold instead is set at $\rm f'_{cut}= 1760 Hz$.
The frequency-related quantities used to characterize the signal are:
\begin{itemize}
\item weighted frequency in the PM region, computed using as weights the pixels energy
%\textcolor{blue}{forse potremmo definire $freq_w$ come $f_w$ poiche' nella formula usiamo $f_{ij}$}

    \begin{equation}
        {\rm freq_w} = \frac{\sum_{ij} {\rm f}_{ij} { \rm en}_{ij}}{\sum_{ij} {\rm en}_{ij}},
    \end{equation}

where ${\rm en}_{ij}$ and ${\rm f}_{ij}$ are, respectively, the energy and central frequency of the pixel $ij$.

\item weighted frequency bandwidth, calculated from the weighted variance: 
 \begin{equation}
{\rm b_w }= \sqrt{\frac{ \sum_{ij} {\rm f}_{ij}^2 {\rm en}_{ij}}{\sum_{ij} {\rm en}_{ij}} -  \left( \frac{\sum_{ij} {\rm f}_{ij} {\rm en}_{ij}}{\sum_{ij} {\rm en}_{ij}} \right) ^2}.
\end{equation}  

\end{itemize}

\begin{figure}[ht!]
\begin{center}
\includegraphics [width = 0.99\textwidth]{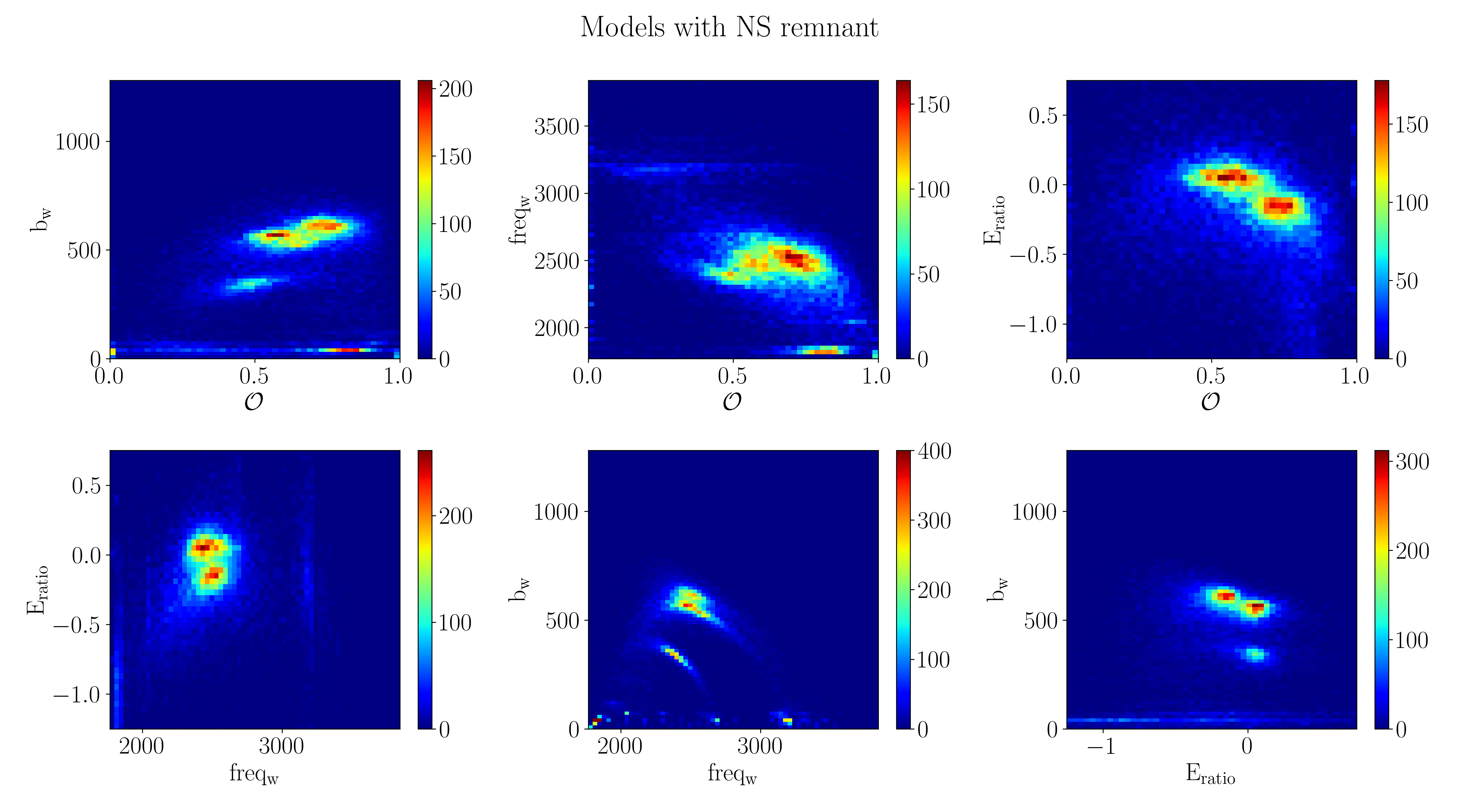}
\caption{\small{Bivariate distributions of the estimated postmerger physical parameters energy ratio $\rm E_{ratio}$, luminosity profile overlap function $\mathcal{O}$, energy-weighted frequency $\rm freq_w$ and bandwidth $\rm b_w$, for simulated waveform models predicting the formation of a NS remnant. Color scales represent the number of counts per bin.%This plot shows waveform signals reconstructed for Livingston the detector.
}}
\label{fig:estimator_dis_pmns}
\end{center}
\end{figure}

\begin{figure}[ht!]
\begin{center}
\includegraphics [width = 0.99\textwidth]{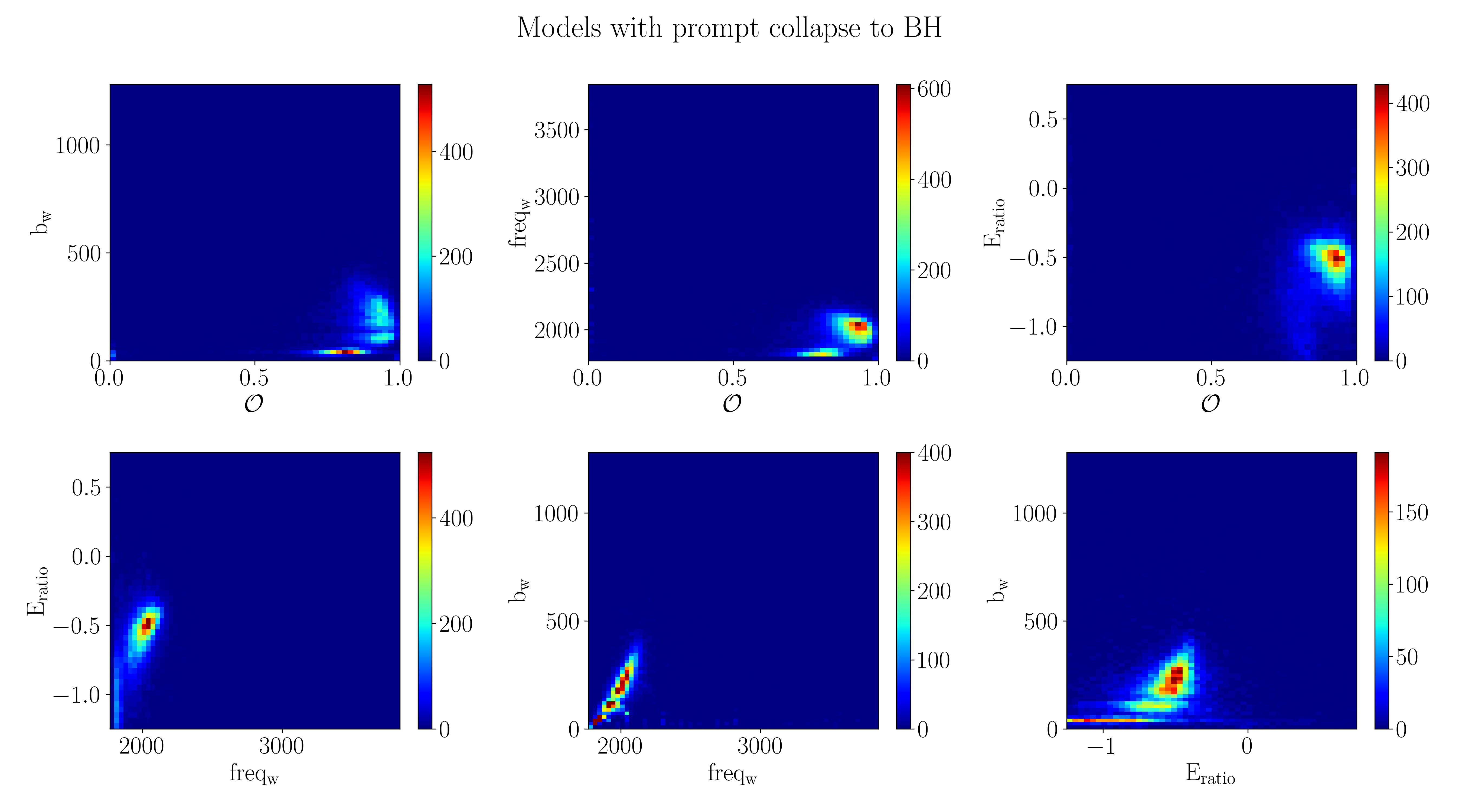}
\caption{\small{Bivariate distributions of the estimated postmerger physical parameters energy ratio $\rm E_{ratio}$, luminosity overlap function $\mathcal{O}$, energy-weighted frequency $\rm freq_w$ and bandwidth $\rm b_w$, for simulated waveform models predicting a prompt collapse to BH scenario. Color scales represent the number of counts per bin. %This plot shows waveform signals reconstructed for Livingston the detector.
}}
\label{fig:estimator_dis_dcbh}
\end{center}
\end{figure}

Among the events simulated as outlined in Sec.~\ref{sec:nr_waveforms}, we consider the ones that satisfy the pre-selection criteria in Eq.~\ref{eq:precut}, and we compute the quantities described above.
%We compute the quantities described above for the events passing the pre-selection cuts among the ones simulated as outlined in Sec.~\ref{sec:nr_waveforms}.
Bivariate distributions of the estimated parameters are shown separately for the two possible scenarios, in Fig.~\ref{fig:estimator_dis_pmns} for the case of BNS coalescence with a massive NS remnant formation and in Fig.~\ref{fig:estimator_dis_dcbh} for models with prompt collapse to BH.

While the distributions referring to the prompt collapse to BH scenario have a compact shape, the NS-remnant scenario shows more complex structures, with clear multi-modality, which is due to the different parameters for the various NR models taken into account. Nonetheless, each NR waveform considered separately results in Gaussian parameter distributions around the expected value.
This suggests that a further development of this analysis can potentially lead to a tool to perform EoS model selection, which is however beyond the goal of this paper.\\


%-------------------------------------------

% RESULTS
%\section{Detection criteria for post-merger emission} 
\section{Criterion to determine the postmerger scenario}
\label{sec:statistics}

We aim to build a criterion to discriminate between the two alternative scenarios following a BNS coalescence. In the following, $H_{0}$ indicates the null hypothesis, i.e. prompt collapse to black hole, while $H_{1}$ stands for presence of a  postmerger GW emission from an excited NS remnant. In our analysis, these alternative hypotheses are represented by the different BNS coalescence models listed in Tab.\ref{tab:NRcatalog}. 

An inspection of the two-dimensional (2D) distributions of the four observable quantities discussed in the previous section (cf.~Fig.~\ref{fig:estimator_dis_pmns} and Fig.~\ref{fig:estimator_dis_dcbh}) indicates that the separation between the two alternative hypotheses appears more evident in some of these 2D projections than others. 
Therefore, the four-dimensional problem can be simplified by selecting the planes $f_1$ $\rm (E_{ratio}, freq_w)$ and $f_2$ $\rm (b_w, \mathcal{O})$, since both appear to have good discriminating power and together they exploit all quantities measured by our method.
This simplification is equivalent to approximating the joint four-dimensional likelihood as a product of separate bivariate likelihoods.

Moreover, $H_1$ distributions show clear multimodal features, each one related to a specific BNS model used in the tuning phase. Because these tuning cases do not cover the entire $H_1$ model parameter space, any detection criterion will need to (i) pass a test based on additional BNS models not used in the tuning, and (ii) avoid being overtuned to such multimodal details of the tuning set. 

\begin{figure}[ht!]
\centering
\subfigure[\label{fig:likez1}]
{\includegraphics [width=0.49\textwidth] {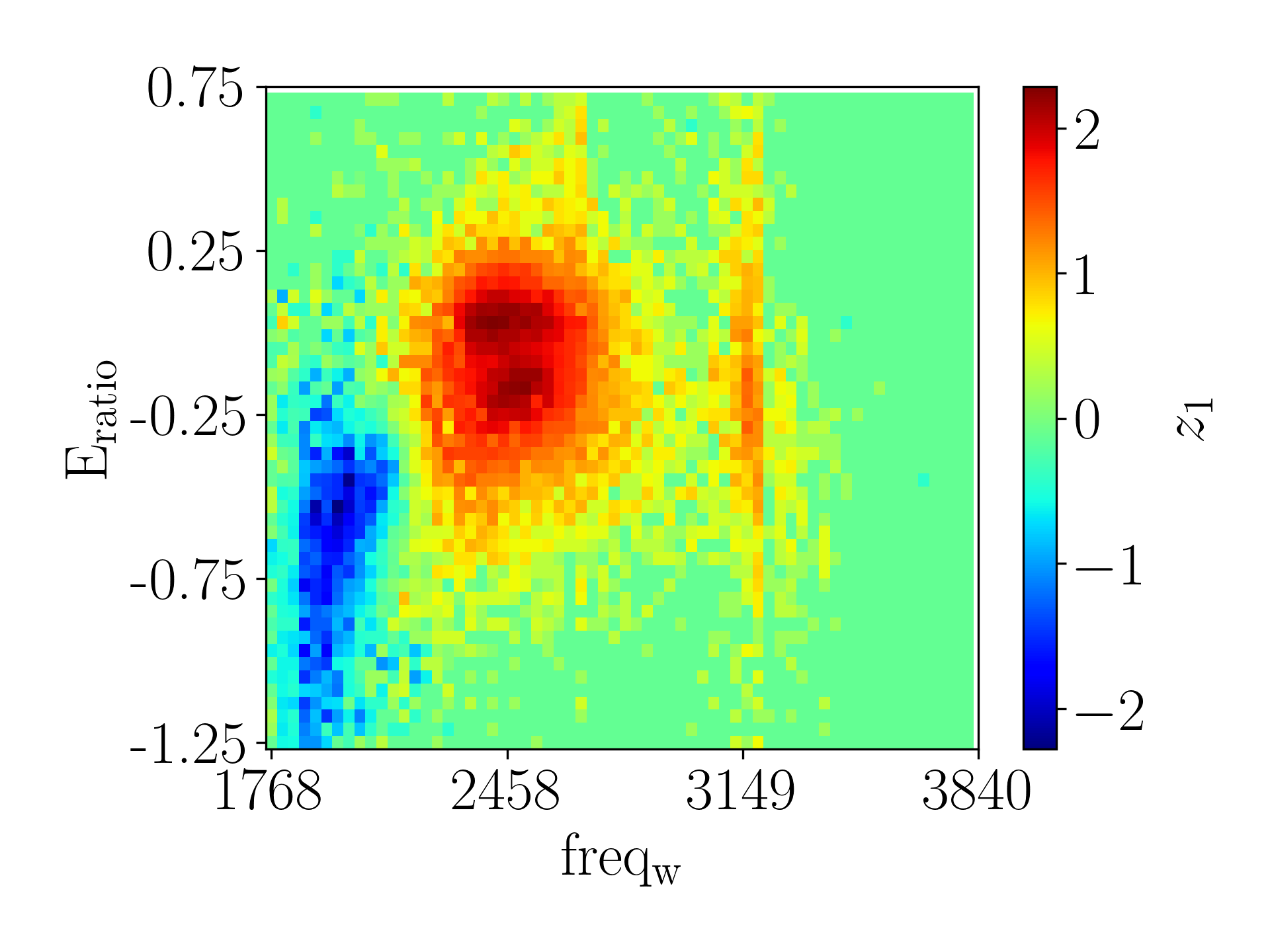}}
\subfigure[\label{fig:likez2}]
{\includegraphics [width=0.49\textwidth]{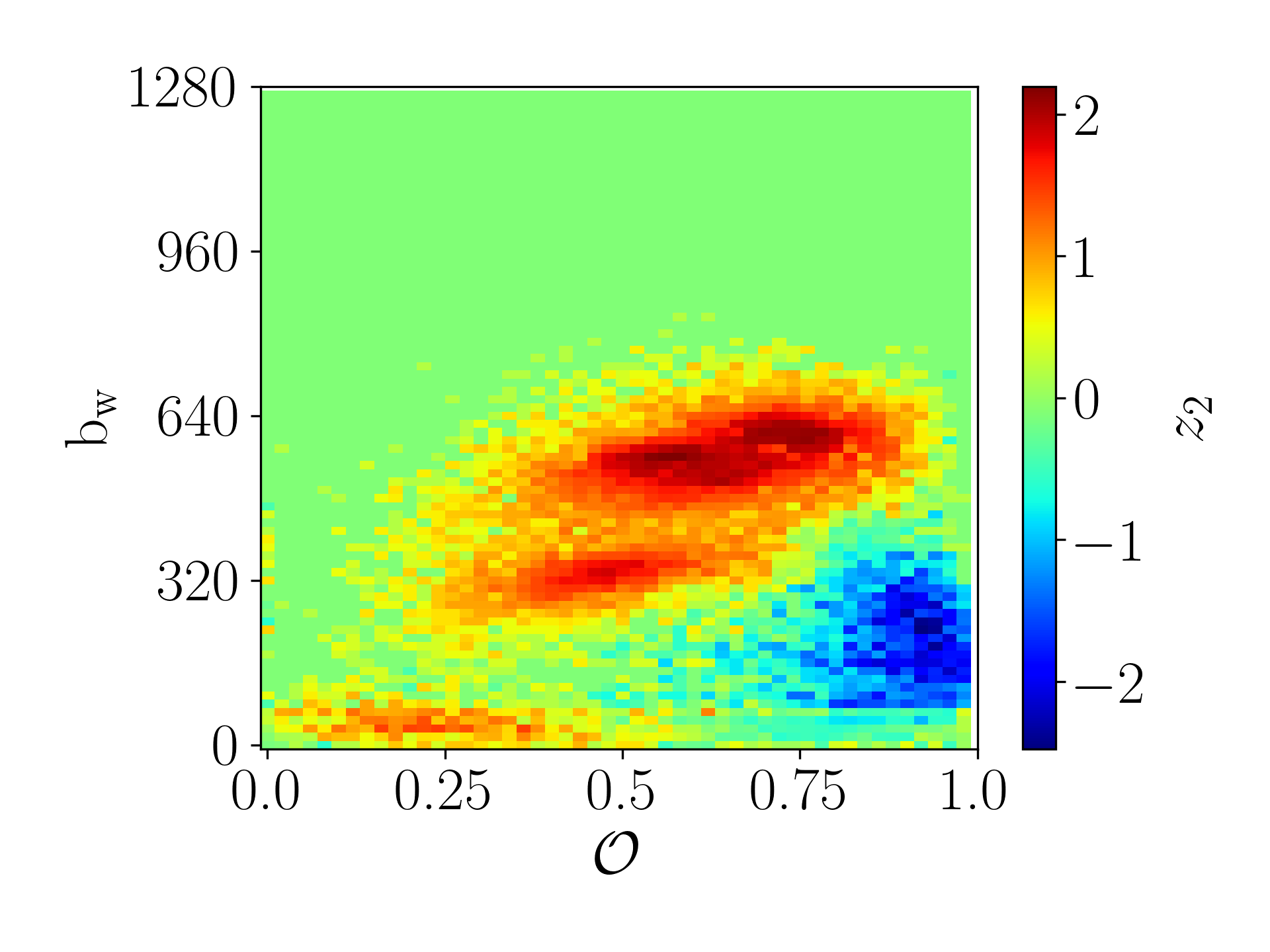}}
\caption{\small{Distributions of the likelihood ratio $z_1$ (left) and $z_2$ (right). Blue regions, corresponding to negative  values of $z_1$ and $z_2$, point to a prompt collapse to BH, while red regions, corresponding to positive values of $z_1$ and $z_2$ point towards the formation of a NS remnant. 
}}
\end{figure}

With these motivations, the proposed detection criterion is based on the likelihood ratios $z_1$ and $z_2$, defined as

\begin{equation}
    z_1 = \log_{10} \left[ \frac{pdf(f_1| H1)}{pdf(f_1| H0)}\right] \hspace{1cm} z_2 = \log_{10} \left[ \frac{pdf(f_2| H1)}{pdf(f_2| H0)}\right]
\end{equation}

where $pdf$ denotes our estimate of the bivariate probability density functions from the empirical histograms appearing in Fig.\ref{fig:estimator_dis_pmns} and \ref{fig:estimator_dis_dcbh}.

In this analysis, we set a lower limit for the minimum value of the $pdf$ estimators corresponding to one single count in each two-dimensional histogram bin. This floor limits the relative statistical uncertainty in the likelihood tails and keeps the likelihood ratios finite. 
Our tuning set includes approximately 90k and 60k simulated events for $H_1$ and $H_0$ respectively. For each hypothesis, the set has been built with the same number of injections for each BNS model, in order to weigh equally the tuning models.
. 

Figure~\ref{fig:likez1} and \ref{fig:likez2} show the distribution of the likelihood ratios $z_1$ and $z_2$ respectively, for the entire set of tuning events that satisfy the pre-selection criteria described by Eq.~\ref{eq:precut}. $z_1$ and $z_2$ assume consistent positive and negative values in the regions of $f_1$ and $f_2$ planes corresponding to the typical parameter values for the two different scenarios.
We also highlight that in some regions of the plotted parameter space, the distributions of $z_1$ and $z_2$ are not informative, because they are fully determined by the ratio of the floor values for the $pdf$ of the two hypotheses.

As anticipated, the two alternative hypotheses appear well separated in the  $z_1$ and $z_2$ values. 
Hence, from the distribution of $z_1$ versus $z_2$ over the set of all tuning events, one can define a scalar statistic, which we call $LDV$ (Likelihood Discrimination Value), for discriminating between the alternative hypotheses

\begin{equation}
    LDV = \log_{10} \left[ \frac{pdf(z_1,z_2 | H1)}{pdf(z_1,z_2|H0)}\right].
    \label{eq:ldv}
\end{equation}

\begin{figure}[ht!]
\centering
\subfigure[\label{fig:likez12}]
{\includegraphics [width=0.47\textwidth] {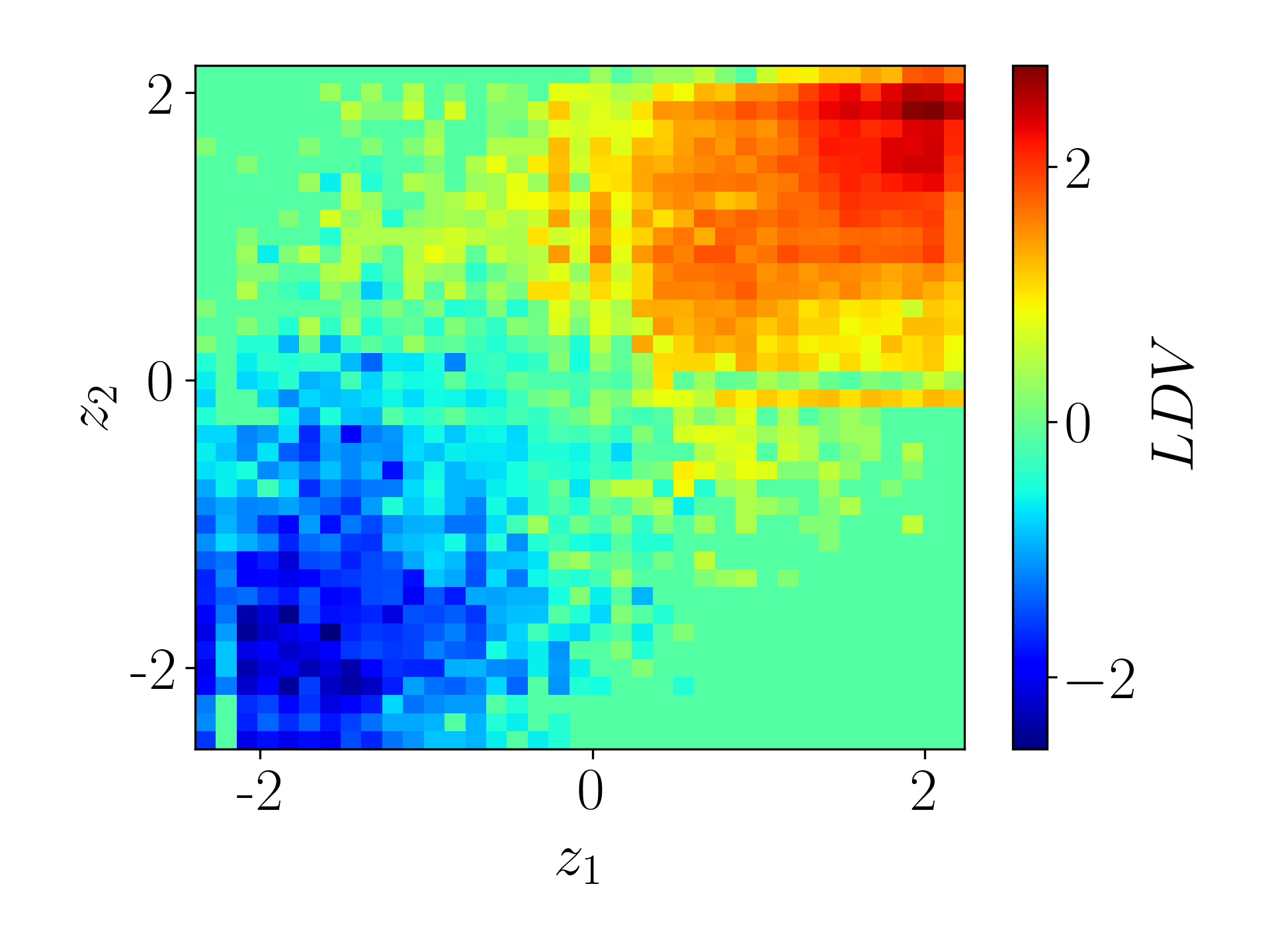}}
\subfigure[\label{fig:likez12_cum}]
{\includegraphics [width=0.47\textwidth]{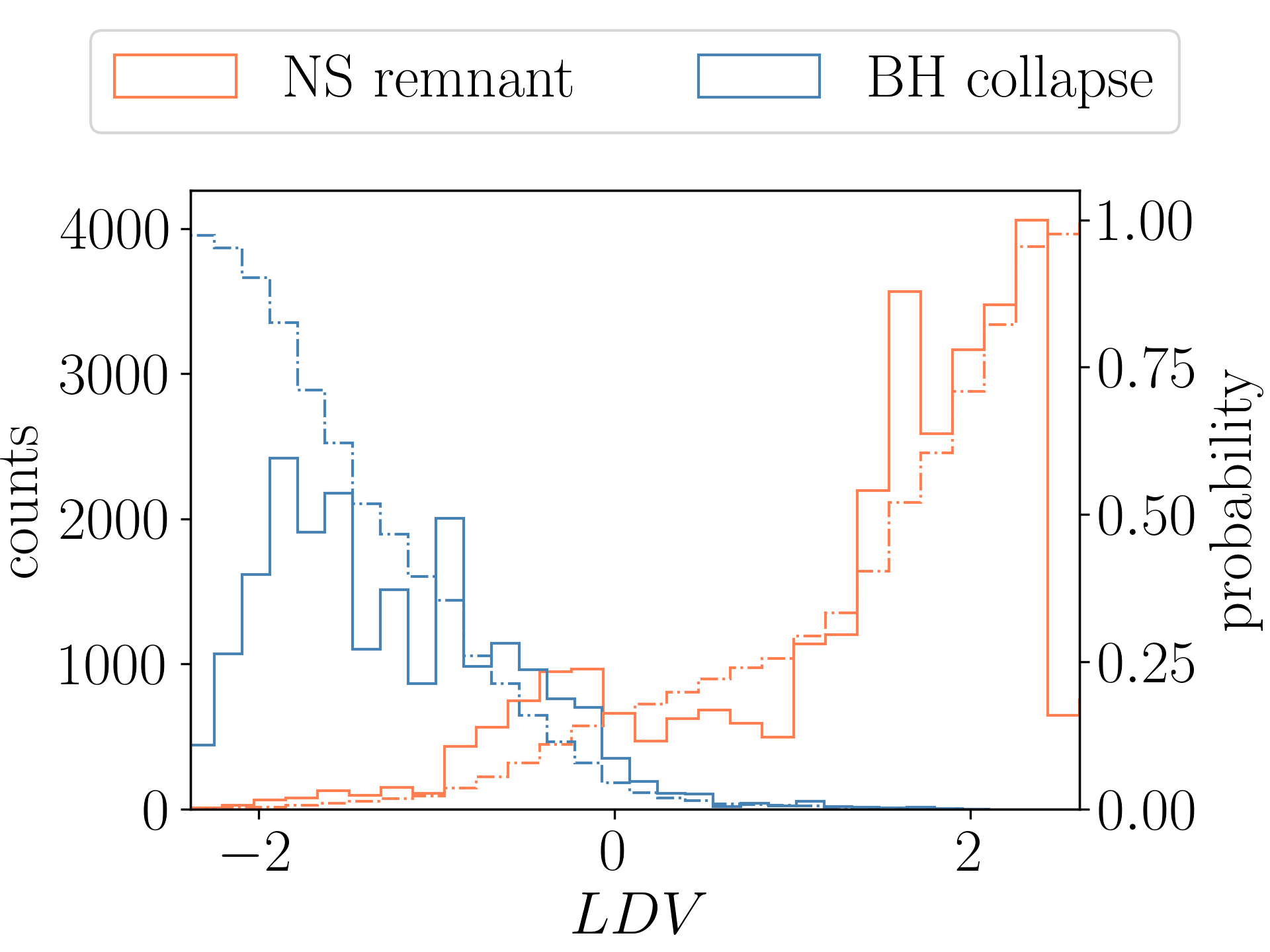}}
\caption{\small{Left panel: $LDV$ distribution over all the simulated events in the tuning set; blue (red) regions, corresponding to negative (positive) values of $LDV$, indicate prompt collapse to BH (formation of a NS remnant). Right panel: Distribution of LDV for the population of events expected to have a prompt collapse to BH in blue, and for those expected to form a massive NS remnant in orange; the dashed-dotted lines show the cumulative distribution of $LDV$ for the events with formation of a NS remnant (in orange), and its inverse cumulative distribution for events with prompt collapse to BH (in blue).}}
\end{figure}

The distribution of $LDV$ over the $(z_1, z_2)$ plane is shown in Fig.\ref{fig:likez12}, while Fig \ref{fig:likez12_cum} shows its cumulative distribution for each of the alternative hypotheses. %of models predicting a prompt or delayed collapse to BH. The 
%se distributions/plots show clearly that the 
%The two regions corresponding to the alternative hypotheses are well separated, and 
The events for the two postmerger scenarios are clustered in different regions of the $(z_1,z_2)$ plane, resulting in specific values of $LDV$; therefore a fixed threshold value on $LDV$ provides a simple criterion to discriminate between them. The same floor limit implemented for the $pdf$ estimators in the $f_1$ and $f_2$ planes has been set also for the $pdf$ estimators in the $z_1$ and $z_2$ plane of Fig.\ref{fig:likez12} and Eq.~\ref{eq:ldv}. In the regions where both $z_1$ and $z_2$ are set to the floor value, we get $LDV_{\rm floor} \simeq -0.1425$, but this value is not informative since fully determined by the chosen floor value. Therefore, when deciding the discriminating threshold value on $LDV$, only values higher than this one should be considered.
%Therefore,  $LDV \simeq -0.1425$ is not informative and only higher threshold values on LDV should be considered.}

\section{Results}
\label{sec:results}
%Once the LDV estimator has been chosen to discern the two scenarios, 

The detection criterion is set by a fixed threshold on $LDV$, defined as explained in the previous Sec.~\ref{sec:statistics}.
%Results have been obtained using as detection criterion a fixed threshold value on $LDV$, defined as explained in the previous Sec.~\ref{sec:statistics}. 
The $LDV$ threshold implicitly sets a mean false alarm probability (FAP), related to the entire set of analyzed $H_0$ signal injections which pass the pre-selection cuts and span a large range of SNR values.  
As usual, the false alarm probability is defined as the fraction of events simulated with a prompt collapse to BH model ($H_0$) that are mis-identified as events consistent with a NS remnant emission ($H_1$). 
In order to keep this FAP of our procedure below $1\%$, we choose a threshold value $LDV_{\rm th} = 0$. In the following, this threshold is used to select candidates for a NS remnant emission.

\begin{figure}[ht!]
\begin{center}
\includegraphics [width=0.7\textwidth]{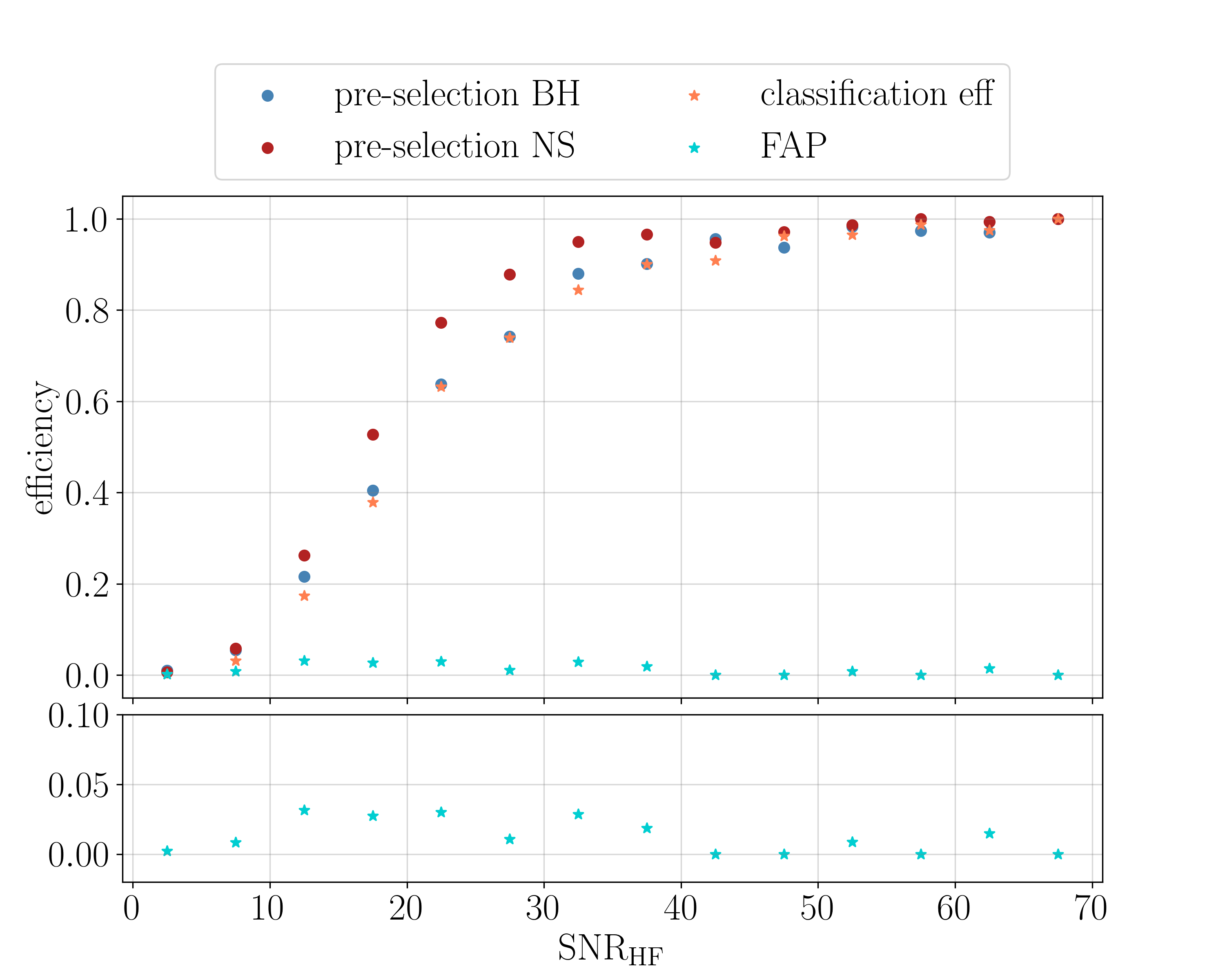}
\caption{\small{Top panel: Efficiencies computed for the sanity check over a subset of roughly $15 \%$ of the full tuning set as a function of the events' $\rm SNR_{HF}$. Dark-blue and red circles represent the percentage of events on which the procedure can be applied, i.e., the events passing the pre-selection criteria, for models predicting a prompt-collapse to BH or a massive NS remnant respectively. Orange stars represent the efficiencies of events correctly recognized as leading to the formation of a NS remnant, while light-blue stars indicate the false alarms. Bottom panel: zoom in the range between 0 and 0.1, to show the fluctuations of FAP values.}}
\label{fig:eff_training}
\end{center}
\end{figure}

A sanity check of the method can be performed by analyzing a random subset of the simulated events used to determine the detection criterion. 
Here we show typical results from a random subset using about 2k simulated events for each BNS model, which corresponds to $\sim 15 \%$ of the full tuning set. 

Figure~\ref{fig:eff_training} reports classification efficiencies and false alarm probabilities as a function of $\rm SNR_{HF}$, which we recall is defined as the signal-to-noise ratio of the simulated BNS event as measured by cWB in the frequency range $768-4096$~Hz. Here, the classification efficiency is defined as the number of events, among all the simulated ones, that are correctly identified to have NS remnant emission in the postmerger.
The classification efficiency is greater than $50 \%$ for injections with $\rm SNR_{HF}>20$, while the false alarm rate is not showing a correlation with $\rm SNR_{HF}$ and fluctuates consistently with its estimated statistical uncertainties, with a mean value of $\sim 1.3\%$. 
Figure~\ref{fig:eff_training} shows that the true positives for a postmerger NS emission approach closely the counts of all simulated events of that class passing our pre-selection criteria. False negatives, or false dismissals, contributed by the detection criterion are therefore limited. For comparison, in the same figure we show also pre-selection efficiencies, i.e., the percentage of events that satisfy the pre-selection criteria, for events that both collapse promptly to a BH and events that form a NS remnant.
Overall, results are self-consistent and the detection criterion passes this sanity check. \\
Lower pre-selection efficiencies at lower $\rm SNR_{HF}$ are mainly due to the fact that a high fraction of simulated events are missed in the pre-selection procedure. However, the pre-selection cuts we chose are functional to the determination of the detection criterion, because taking into account events with a too low postmerger signal would add just noise in the distributions of the parameters that we exploited in the tuning phase of the analysis, described in Sec.~\ref{sec:statistics}. \\
Relaxing the pre-selection criteria to less stringent constraints just for the final analysis would be possible, but we expect that the efficiencies of the correct classification as NS remnant would be practically unchanged. In fact, the number of analyzed events with enough signal in the postmerger region would be the same, and the current classification criterion would perform the same for events with $\rm SNR_{rec} > 10 $. %, and by keeping fixed the current detection criteria the NS remnant  correctly identified as is done here. 
Instead, for events with lower $\rm SNR_{rec}$, the resulting increase in pre-selection efficiency %which is expected to increase, because we will have more events passing the less stringent pre-selection cuts, 
would bring in more events dominated by noise fluctuations, therefore pushing towards a balance of the counts of false alarms and true positives, and therefore damaging the purity of our detections in the low SNR range.   %since the events that lead to the formation of a NS remnant, but have not sufficient signal in the postmerger region, will be misidentified as promptly collapsing to BH. 

\begin{figure}[ht!]
\begin{center}
\includegraphics [width=0.7\textwidth]{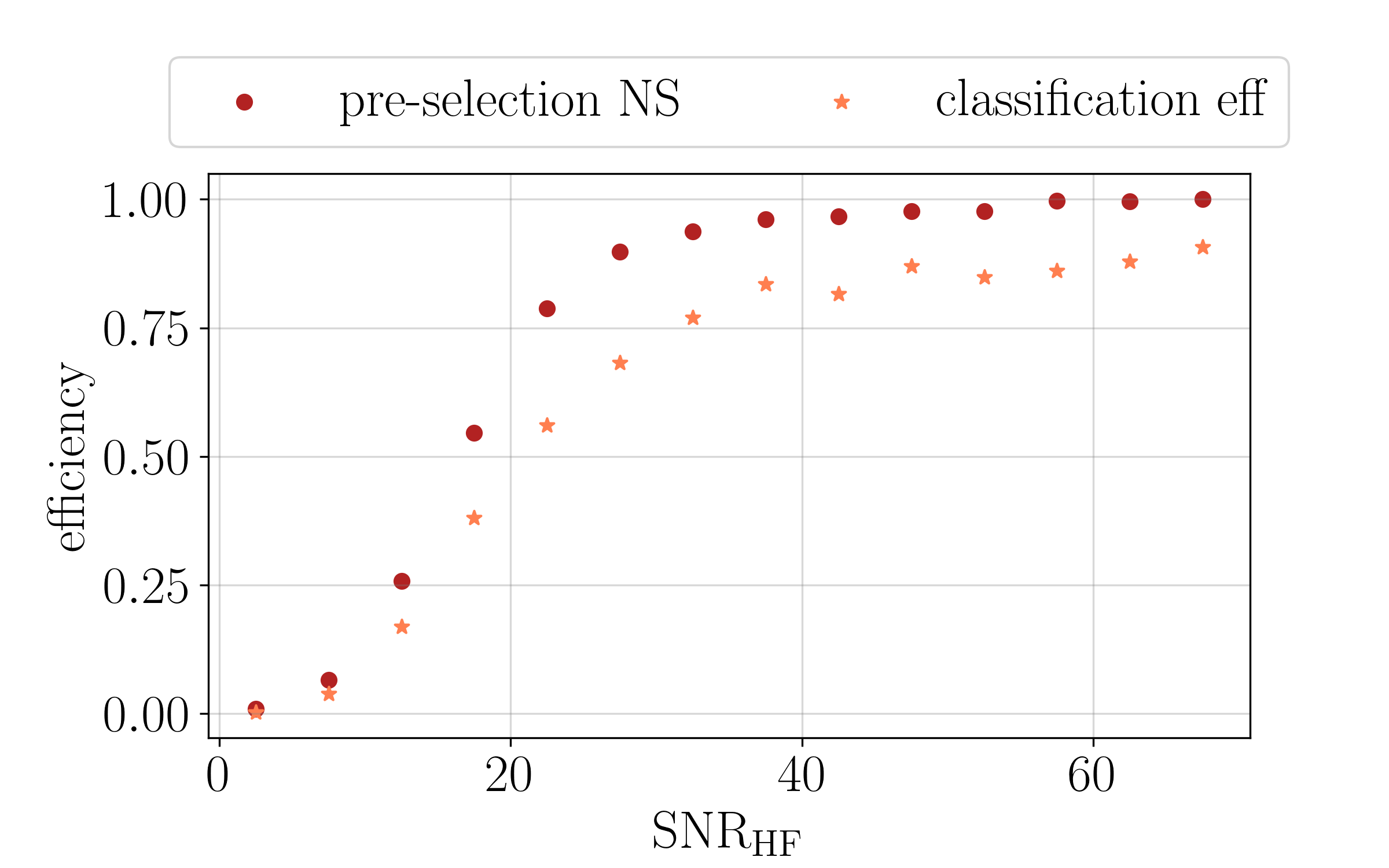}
%{figure/eff_snrhf_newwaveforms_allevents.png}
\caption{\small{Example of efficiencies for a new set of simulated events (generated with different waveform models, as explained in the text), shown as a function of $\rm SNR_{HF}$. }}
\label{fig:eff_new}
\end{center}
\end{figure}

\begin{figure}[ht!]
\begin{center}
\includegraphics [width=0.7\textwidth]{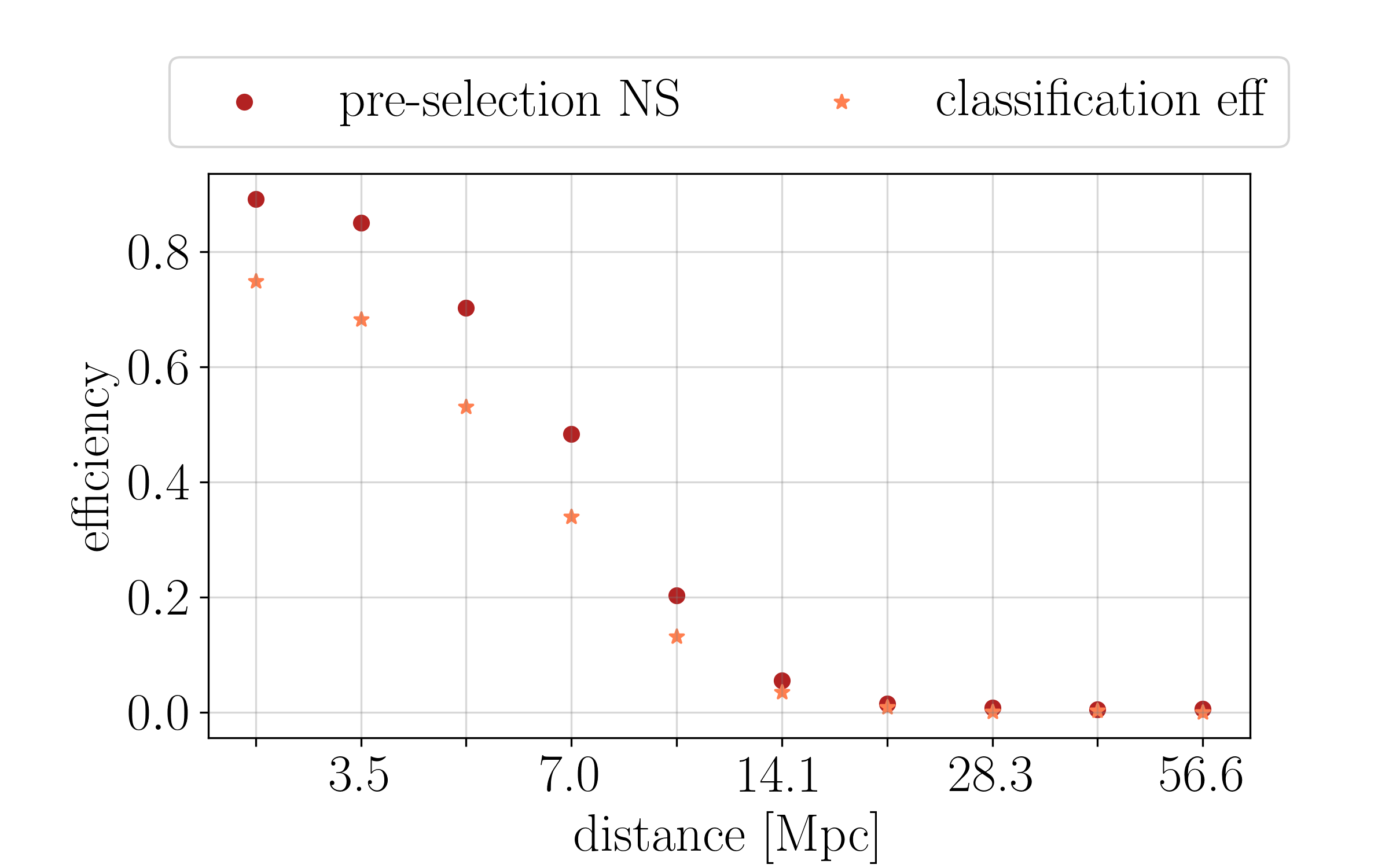}
%{figure/eff_dist_correctorder.png}
\caption{\small{Efficiencies of our method, on a new set of simulated events, as a function of the source distance in Mpc. }}
\label{fig:eff_new_dist}
\end{center}
\end{figure}

%\subsection{results on a novel set of events}
Finally, we test the performance of the method on a fully different set of events simulated from three BNS models not included in the development and tuning previously described. These models are listed in Tab.\ref{tab:NRcatalog}, where they can be identified by the name in bold, and have postmerger emission from a NS remnant, hence pertaining to hypothesis $H_1$. About 3k events have been analyzed for each BNS model. 

Resulting classification efficiency and false alarm probability of the method are shown in Fig.~\ref{fig:eff_new} as a function of $\rm SNR_{HF}$. 
Performances are consistent with those achieved on the subsample of the tuning set within statistical uncertainties, showing that the method works as well for a class of BNS models wider than those included in the tuning set. 
The same considerations on the stability of the performances with respect to pre-selection criteria hold here as well.

For a better interpretation of the previous results %(see Fig.~\ref{fig:eff_new}) 
from the astrophysical point of view, Fig. \ref{fig:eff_new_dist} shows efficiency and false alarm probability as a function of the simulated source distance. 
At Advanced LIGO and Advanced Virgo design sensitivities, this method shows interesting efficiencies only up to the closest edge of the Virgo Cluster.
Considering that the estimated luminosity distance for GW170817 was roughly 40 Mpc, also if a postmerger signal were present, our method would not have detected it, since efficiencies drop below $1\%$ above 20 Mpc.
%it would not have passed our pre-selection criteria. 
\mycomment{
Assuming a uniform distribution in comoving volume of BNS mergers, the effective volume of post-merger emission by NS remnants which is surveyed by this method is %\textcolor{red}{plot integral = 7.05603}
\begin{equation}
    V = 4 \pi \int_0^\infty dr \ r^2 \epsilon (r) \sim 7 Mpc^3,
\end{equation}

where $\epsilon (r)$ is the measured classification efficiency as a function of source distance.
}

\mycomment{
As these results show, currently this method is not applicable to the BNS merger signals detected up to now.
%astrophysical sources, since, also if they emit a postmerger GW signal, it will not pass our pre-selection criteria.
Further developments of this analysis, in which we investigate the possibility of less restraining pre-selection criteria, and we possibly implement the techniques described in \cite{Miani:2023mgl} to enhance the detection of weak signals, will increase the effective detection volume of our procedure.
Together with the increased detector sensitivities and improvements of the underlying cWB pipeline, this will allow us to actually observe astrophysical populations of BNS systems in the future.
}

\section{Conclusions}
\label{sec:conclusions}
% CONCLUSION
We presented a new method to characterize the postmerger GW emission from BNS systems, and to discriminate between the two possible scenarios: the formation of a NS remnant or prompt collapse to black hole after the merger. The analysis is developed in the framework of the cWB algorithm and is completely model- and morphology-independent. Based on the TF map of the reconstructed signal, we define four quantities to describe the physical properties of the signal: ratio between energy in the postmerger and in the late-merger region; luminosity profile overlap function; energy weighted frequency; and energy weighted bandwidth. We computed these quantities for a set of simulated events obtained by injections of NR waveforms, taking into consideration only events that pass some pre-selection criteria that ensure we analyze only triggers with some reconstructed signal energy after the merger. The models we consider include both the case in which the system collapses to a BH immediately after the merger, and the case in which it forms an excited NS remnant that survives for some time before collapsing to a BH and thus emits a GW postmerger signal. We find that for events simulated with models belonging to these different scenarios, the values of the parameters listed above follow very different distributions. We therefore use the bivariate distributions of these parameters to define a statistic that allows us to discriminate between prompt collapse to BH, or formation of a NS remnant. We test the performance of our procedure, computing its efficiency and false alarm probability, on a new set of events generated with NR waveforms different from the ones included in the tuning. We find that our method yields a high efficiency for the events that pass our pre-selection criteria, and the false alarm probability of the search for a NS remnant postmerger signal is kept under control at $ \le 1\%$.

The overall efficiency declines fast with increasing source distance, and GW170817 is still outside our detection range. 
%currently our procedure cannot target astrophysical BNS sources, and also an event like GW170817 would have been outside our detection range.
Further developments of this analysis, in which we investigate the possibility of less restraining pre-selection criteria, and we possibly implement the techniques described in \cite{Miani:2023mgl} to enhance the detection of weak signals, will increase the effective detection volume of our procedure.
Together with the increased detector sensitivities and improvements of the underlying cWB pipeline, this will allow us to actually observe astrophysical populations of BNS systems in the future.
%There is an ongoing effort to increase cWB reconstruction capability for weak signals, and adapting such improvements to our procedure will boost our detection capability and allow us to employ less restraining pre-selection cuts. 
This analysis is based on the coherent reconstruction of a GW candidate in the detector network provided by cWB, however here we exploited only the projection of this information on the most sensitive detector, Livingston. Therefore, further improvement is expected by exploiting the full information on the entire network.
Finally, detectors' technological improvements will enhance their sensitivity also in the kilohertz region, increasing the probability of a postmerger detection. Together with the improvements planned for Advanced Virgo+ \cite{Flaminio} and Advanced LIGO+ \cite{AdLIGO2022}, also third-generation detectors will come into play to make postmerger detections feasible. For example, Einstein Telescope, with its xylophone configuration \cite{Hild:2010id, Punturo2010,ET2020}, is expected to increase the sensitivity in the high-frequency range by at least a factor 10 with respect to current detectors \cite{KAGRA:2013rdx}, which means that sources at roughly 70 Mpc will be a feasible target for our analysis.\\

\section*{Acknowledgments}

We thank Wolfgang Kastaun and Francesco Salemi for the useful discussion. The authors are grateful for computational resources provided by the LIGO Laboratory and supported by the National Science Foundation Grants No.~PHY-0757058 and No.~PHY-0823459. This research has made use of data, software and/or web tools obtained from the Gravitational Wave Open Science Center (https://www.gw-openscience.org), a service of LIGO Laboratory, the 
LIGO Scientific Collaboration and the Virgo Collaboration. LIGO is funded by 
the U.S. National Science Foundation. Virgo is funded by the French Centre 
National de Recherche Scientifique (CNRS), the Italian Istituto Nazionale della 
Fisica Nucleare (INFN) and the Dutch Nikhef, with contributions by Polish and Hungarian institutes.
A.P.~is supported by the research programme 
of the Netherlands Organisation for Scientific Research (NWO).

% Bibliography
\clearpage
\newpage
\bibliographystyle{unsrt}
\bibliography{reference}

\begin{thebibliography}{10}

\bibitem{Advanced_LIGO}
The LIGO~Scientific Collaboration.
\newblock {Advanced LIGO}.
\newblock {\em Classical and Quantum Gravity}, 32(7):074001, 2015.

\bibitem{LIGOScientific:2014pky}
J.~Aasi et~al.
\newblock {Advanced LIGO}.
\newblock {\em Class. Quant. Grav.}, 32:074001, 2015.

\bibitem{Advanced_Virgo}
Ilaria Nardecchia.
\newblock {Detecting Gravitational Waves with Advanced Virgo}.
\newblock {\em Galaxies}, 10(1), 2022.

\bibitem{VIRGO:2014yos}
F.~Acernese et~al.
\newblock {Advanced Virgo: a second-generation interferometric gravitational
  wave detector}.
\newblock {\em Class. Quant. Grav.}, 32(2):024001, 2015.

\bibitem{GWTC-1}
B.~P. et~al. Abbott.
\newblock {GWTC-1: A Gravitational-Wave Transient Catalog of Compact Binary
  Mergers Observed by LIGO and Virgo during the First and Second Observing
  Runs}.
\newblock {\em Phys. Rev. X}, 9:031040, Sep 2019.

\bibitem{Abbott_2020niy}
LIGO~Scientific Collaboration and Virgo Collaboration.
\newblock {GWTC-2: Compact Binary Coalescences Observed by LIGO and Virgo
  during the First Half of the Third Observing Run}.
\newblock {\em Phys. Rev. X}, 11:021053, Jun 2021.

\bibitem{GWTC-3}
{The LIGO Scientific Collaboration, the Virgo Collaboration, the KAGRA
  Collaboration}.
\newblock {GWTC-3: Compact Binary Coalescences Observed by LIGO and Virgo
  During the Second Part of the Third Observing Run}.
\newblock {\url{https://arxiv.org/abs/2111.03606v2}}.

\bibitem{TheLIGOScientific:2017qsa}
B.~P. Abbott et~al.
\newblock {GW170817: Observation of Gravitational Waves from a Binary Neutron
  Star Inspiral}.
\newblock {\em Phys. Rev. Lett.}, 119(16):161101, 2017.

\bibitem{Abbott:2020uma}
B.~P. Abbott et~al.
\newblock {GW190425: Observation of a Compact Binary Coalescence with Total
  Mass $\sim 3.4 M_{\odot}$}.
\newblock {\em Astrophys. J. Lett.}, 892(1):L3, 2020.

\bibitem{LIGOScientific:2017ync}
B.~P. Abbott et~al.
\newblock {Multi-messenger Observations of a Binary Neutron Star Merger}.
\newblock {\em Astrophys. J. Lett.}, 848(2):L12, 2017.

\bibitem{Dietrich:2020eud}
Tim Dietrich, Tanja Hinderer, and Anuradha Samajdar.
\newblock {Interpreting Binary Neutron Star Mergers: Describing the Binary
  Neutron Star Dynamics, Modelling Gravitational Waveforms, and Analyzing
  Detections}.
\newblock {\em Gen. Rel. Grav.}, 53(3):27, 2021.

\bibitem{Hinderer:2009ca}
Tanja Hinderer, Benjamin~D. Lackey, Ryan~N. Lang, and Jocelyn~S. Read.
\newblock {Tidal deformability of neutron stars with realistic equations of
  state and their gravitational wave signatures in binary inspiral}.
\newblock {\em Phys. Rev.}, D81:123016, 2010.

\bibitem{Damour:2012yf}
Thibault Damour, Alessandro Nagar, and Loic Villain.
\newblock {Measurability of the tidal polarizability of neutron stars in
  late-inspiral gravitational-wave signals}.
\newblock {\em Phys. Rev.}, D85:123007, 2012.

\bibitem{DelPozzo:2013ala}
Walter Del~Pozzo, Tjonnie G.~F. Li, Michalis Agathos, Chris Van Den~Broeck, and
  Salvatore Vitale.
\newblock {Demonstrating the feasibility of probing the neutron star equation
  of state with second-generation gravitational wave detectors}.
\newblock {\em Phys. Rev. Lett.}, 111(7):071101, 2013.

\bibitem{Lackey:2014fwa}
Benjamin~D. Lackey and Leslie Wade.
\newblock {Reconstructing the neutron-star equation of state with
  gravitational-wave detectors from a realistic population of inspiralling
  binary neutron stars}.
\newblock {\em Phys. Rev.}, D91(4):043002, 2015.

\bibitem{Agathos:2015uaa}
Michalis Agathos, Jeroen Meidam, Walter Del~Pozzo, Tjonnie G.~F. Li, Marco
  Tompitak, John Veitch, Salvatore Vitale, and Chris Van Den~Broeck.
\newblock {Constraining the neutron star equation of state with gravitational
  wave signals from coalescing binary neutron stars}.
\newblock {\em Phys. Rev.}, D92(2):023012, 2015.

\bibitem{Dietrich:2018uni}
Tim Dietrich et~al.
\newblock {Matter imprints in waveform models for neutron star binaries: Tidal
  and self-spin effects}.
\newblock {\em Phys. Rev.}, D99(2):024029, 2019.

\bibitem{LIGOScientific:2018hze}
B.~P. Abbott et~al.
\newblock {Properties of the binary neutron star merger GW170817}.
\newblock {\em Phys. Rev. X}, 9(1):011001, 2019.

\bibitem{Piro:2017zec}
Anthony~L. Piro, Bruno Giacomazzo, and Rosalba Perna.
\newblock {The Fate of Neutron Star Binary Mergers}.
\newblock {\em Astrophys. J. Lett.}, 844(2):L19, 2017.

\bibitem{Sarin:2020gxb}
Nikhil Sarin and Paul~D. Lasky.
\newblock {The evolution of binary neutron star post-merger remnants: a
  review}.
\newblock {\em Gen. Rel. Grav.}, 53(6):59, 2021.

\bibitem{LIGOScientific:2017fdd}
B.~P. Abbott et~al.
\newblock {Search for Post-merger Gravitational Waves from the Remnant of the
  Binary Neutron Star Merger GW170817}.
\newblock {\em Astrophys. J. Lett.}, 851(1):L16, 2017.

\bibitem{LIGOScientific:2018urg}
B.~P. Abbott et~al.
\newblock {Search for gravitational waves from a long-lived remnant of the
  binary neutron star merger GW170817}.
\newblock {\em Astrophys. J.}, 875(2):160, 2019.

\bibitem{Virgosqueezing}
The~Virgo Collaboration.
\newblock {Increasing the Astrophysical Reach of the Advanced Virgo Detector
  via the Application of Squeezed Vacuum States of Light}.
\newblock {\em Phys. Rev. Lett.}, 123:231108, Dec 2019.

\bibitem{LIGOsqueezing}
The~LIGO Collaboration.
\newblock {Quantum-Enhanced Advanced LIGO Detectors in the Era of
  Gravitational-Wave Astronomy}.
\newblock {\em Phys. Rev. Lett.}, 123:231107, Dec 2019.

\bibitem{Punturo:2010zz}
M.~Punturo et~al.
\newblock {The Einstein Telescope: A third-generation gravitational wave
  observatory}.
\newblock {\em Class. Quant. Grav.}, 27:194002, 2010.

\bibitem{Maggiore:2019uih}
Michele Maggiore et~al.
\newblock {Science Case for the Einstein Telescope}.
\newblock {\em JCAP}, 03:050, 2020.

\bibitem{Freise:2008dk}
A.~Freise, S.~Chelkowski, S.~Hild, W.~Del~Pozzo, A.~Perreca, and A.~Vecchio.
\newblock {Triple Michelson Interferometer for a Third-Generation Gravitational
  Wave Detector}.
\newblock {\em Class. Quant. Grav.}, 26:085012, 2009.

\bibitem{Hild:2009ns}
Stefan Hild, Simon Chelkowski, Andreas Freise, Janyce Franc, Nazario Morgado,
  Raffaele Flaminio, and Riccardo DeSalvo.
\newblock {A Xylophone Configuration for a third Generation Gravitational Wave
  Detector}.
\newblock {\em Class. Quant. Grav.}, 27:015003, 2010.

\bibitem{Sathyaprakash:2011bh}
B.~Sathyaprakash et~al.
\newblock {Scientific Potential of Einstein Telescope}.
\newblock In {\em {46th Rencontres de Moriond on Gravitational Waves and
  Experimental Gravity}}, pages 127--136, 8 2011.

\bibitem{Reitze:2019iox}
David Reitze et~al.
\newblock {Cosmic Explorer: The U.S. Contribution to Gravitational-Wave
  Astronomy beyond LIGO}.
\newblock {\em Bull. Am. Astron. Soc.}, 51(7):035, 2019.

\bibitem{Evans:2021gyd}
Matthew Evans et~al.
\newblock {A Horizon Study for Cosmic Explorer: Science, Observatories, and
  Community}.
\newblock 9 2021.

\bibitem{Clark:2014wua}
J.~Clark, A.~Bauswein, L.~Cadonati, H.~T. Janka, C.~Pankow, and N.~Stergioulas.
\newblock {Prospects For High Frequency Burst Searches Following Binary Neutron
  Star Coalescence With Advanced Gravitational Wave Detectors}.
\newblock {\em Phys. Rev.}, D90(6):062004, 2014.

\bibitem{Clark:2015zxa}
James~Alexander Clark, Andreas Bauswein, Nikolaos Stergioulas, and Deirdre
  Shoemaker.
\newblock {Observing Gravitational Waves From The Post-Merger Phase Of Binary
  Neutron Star Coalescence}.
\newblock {\em Class. Quant. Grav.}, 33(8):085003, 2016.

\bibitem{Chatziioannou:2017ixj}
Katerina Chatziioannou, James~Alexander Clark, Andreas Bauswein, Margaret
  Millhouse, Tyson~B. Littenberg, and Neil Cornish.
\newblock {Inferring the post-merger gravitational wave emission from binary
  neutron star coalescences}.
\newblock {\em Phys. Rev. D}, 96(12):124035, 2017.

\bibitem{Bauswein:2011tp}
A.~Bauswein and H.~Th. Janka.
\newblock {Measuring neutron-star properties via gravitational waves from
  binary mergers}.
\newblock {\em Phys. Rev. Lett.}, 108:011101, 2012.

\bibitem{Takami:2014zpa}
Kentaro Takami, Luciano Rezzolla, and Luca Baiotti.
\newblock {Constraining the Equation of State of Neutron Stars from Binary
  Mergers}.
\newblock {\em Phys. Rev. Lett.}, 113(9):091104, 2014.

\bibitem{Rezzolla:2016nxn}
Luciano Rezzolla and Kentaro Takami.
\newblock {Gravitational-wave signal from binary neutron stars: a systematic
  analysis of the spectral properties}.
\newblock {\em Phys. Rev.}, D93(12):124051, 2016.

\bibitem{Bernuzzi:2015rla}
Sebastiano Bernuzzi, Tim Dietrich, and Alessandro Nagar.
\newblock {Modeling the complete gravitational wave spectrum of neutron star
  mergers}.
\newblock {\em Phys. Rev. Lett.}, 115(9):091101, 2015.

\bibitem{Bauswein:2012ya}
A.~Bauswein, H.~T. Janka, K.~Hebeler, and A.~Schwenk.
\newblock {Equation-of-state dependence of the gravitational-wave signal from
  the ring-down phase of neutron-star mergers}.
\newblock {\em Phys. Rev.}, D86:063001, 2012.

\bibitem{Hotokezaka:2013iia}
Kenta Hotokezaka, Kenta Kiuchi, Koutarou Kyutoku, Takayuki Muranushi, Yu-ichiro
  Sekiguchi, Masaru Shibata, and Keisuke Taniguchi.
\newblock {Remnant massive neutron stars of binary neutron star mergers:
  Evolution process and gravitational waveform}.
\newblock {\em Phys. Rev.}, D88:044026, 2013.

\bibitem{Bauswein:2014qla}
A.~Bauswein, N.~Stergioulas, and H.~T. Janka.
\newblock {Revealing the high-density equation of state through binary neutron
  star mergers}.
\newblock {\em Phys. Rev.}, D90(2):023002, 2014.

\bibitem{Takami:2014tva}
Kentaro Takami, Luciano Rezzolla, and Luca Baiotti.
\newblock {Spectral properties of the post-merger gravitational-wave signal
  from binary neutron stars}.
\newblock {\em Phys. Rev.}, D91(6):064001, 2015.

\bibitem{Bauswein:2015yca}
A.~Bauswein and N.~Stergioulas.
\newblock {Unified picture of the post-merger dynamics and gravitational wave
  emission in neutron star mergers}.
\newblock {\em Phys. Rev.}, D91(12):124056, 2015.

\bibitem{Lioutas:2021jbl}
Georgios Lioutas, Andreas Bauswein, and Nikolaos Stergioulas.
\newblock {Frequency deviations in universal relations of isolated neutron
  stars and postmerger remnants}.
\newblock {\em Phys. Rev. D}, 104(4):043011, 2021.

\bibitem{Breschi:2019srl}
Matteo Breschi, Sebastiano Bernuzzi, Francesco Zappa, Michalis Agathos, Albino
  Perego, David Radice, and Alessandro Nagar.
\newblock {kiloHertz gravitational waves from binary neutron star remnants:
  time-domain model and constraints on extreme matter}.
\newblock {\em Phys. Rev. D}, 100(10):104029, 2019.

\bibitem{Easter:2020ifj}
Paul~J. Easter, Sudarshan Ghonge, Paul~D. Lasky, Andrew~R. Casey, James~A.
  Clark, Francisco~Hernandez Vivanco, and Katerina Chatziioannou.
\newblock {Detection and parameter estimation of binary neutron star merger
  remnants}.
\newblock {\em Phys. Rev. D}, 102(4):043011, 2020.

\bibitem{Soultanis:2021oia}
Theodoros Soultanis, Andreas Bauswein, and Nikolaos Stergioulas.
\newblock {Analytic models of the spectral properties of gravitational waves
  from neutron star merger remnants}.
\newblock {\em Phys. Rev. D}, 105(4):043020, 2022.

\bibitem{Breschi:2022xnc}
Matteo Breschi, Sebastiano Bernuzzi, Kabir Chakravarti, Alessandro Camilletti,
  Aviral Prakash, and Albino Perego.
\newblock {Kilohertz Gravitational Waves From Binary Neutron Star Mergers:
  Numerical-relativity Informed Postmerger Model}.
\newblock 5 2022.

\bibitem{Puecher:2022oiz}
Anna Puecher, Tim Dietrich, Ka~Wa Tsang, Chinmay Kalaghatgi, Soumen Roy,
  Yoshinta Setyawati, and Chris Van Den~Broeck.
\newblock {Unraveling information about supranuclear-dense matter from the
  complete binary neutron star coalescence process using future
  gravitational-wave detector networks}.
\newblock 10 2022.

\bibitem{Wijngaarden:2022sah}
Marcella Wijngaarden, Katerina Chatziioannou, Andreas Bauswein, James~A. Clark,
  and Neil~J. Cornish.
\newblock {Probing neutron stars with the full premerger and postmerger
  gravitational wave signal from binary coalescences}.
\newblock {\em Phys. Rev. D}, 105(10):104019, 2022.

\bibitem{Easter:2018pqy}
Paul~J. Easter, Paul~D. Lasky, Andrew~R. Casey, Luciano Rezzolla, and Kentaro
  Takami.
\newblock {Computing Fast and Reliable Gravitational Waveforms of Binary
  Neutron Star Merger Remnants}.
\newblock {\em Phys. Rev. D}, 100(4):043005, 2019.

\bibitem{Klimenko:2015ypf}
S.~Klimenko et~al.
\newblock {Method for detection and reconstruction of gravitational wave
  transients with networks of advanced detectors}.
\newblock {\em Phys. Rev. D}, 93(4):042004, 2016.

\bibitem{Klimenko:2008fu}
S.~Klimenko, I.~Yakushin, A.~Mercer, and Guenakh Mitselmakher.
\newblock {Coherent method for detection of gravitational wave bursts}.
\newblock {\em Class. Quant. Grav.}, 25:114029, 2008.

\bibitem{Klimenko:2005xv}
S.~Klimenko, S.~Mohanty, Malik Rakhmanov, and Guenakh Mitselmakher.
\newblock {Constraint likelihood analysis for a network of gravitational wave
  detectors}.
\newblock {\em Phys. Rev. D}, 72:122002, 2005.

\bibitem{klimenko_sergey_2021_4419902}
Sergey Klimenko, Gabriele Vedovato, Valentin Necula, Francesco Salemi, Marco
  Drago, Eric Chassande-Mottin, Vaibhav Tiwari, Claudia Lazzaro, Brendan
  O'Brian, Marek Szczepanczyk, Shubhanshu Tiwari, and V.~Gayathri.
\newblock cwb pipeline library: 6.4.0, January 2021.

\bibitem{Drago:2020kic}
M.~Drago et~al.
\newblock {Coherent WaveBurst, a pipeline for unmodeled gravitational-wave data
  analysis}.
\newblock 6 2020.

\bibitem{Kastaun15}
Wolfgang Kastaun and Filippo Galeazzi.
\newblock Properties of hypermassive neutron stars formed in mergers of
  spinning binaries.
\newblock {\em Phys. Rev. D}, 91:064027, Mar 2015.

\bibitem{Endrizzi16}
A.Endrizzi, R.~Ciolfi, B.~Giacomazzo, W.~Kastaun, and T.~Kawamura.
\newblock {\it General relativistic magnetohydrodynamic simulations of binary
  neutron star mergers with the APR4 equation of state}.
\newblock {\em Classical and Quantum Gravity}, 33(16):164001, 2016.

\bibitem{Kawamura16}
T.~Kawamura, B.~Giacomazzo, W.~Kastaun, R.~Ciolfi, A.~Endrizzi, L.~Baiotti, and
  R.~Perna.
\newblock {\it Binary neutron star mergers and short gamma-ray bursts: Effects
  of magnetic field orientation, equation of state, and mass ratio}.
\newblock {\em Phys. Rev. D}, 94:064012, Sep 2016.

\bibitem{Endrizzi18}
Andrea Endrizzi, Domenico Logoteta, Bruno Giacomazzo, Ignazio Bombaci, Wolfgang
  Kastaun, and Riccardo Ciolfi.
\newblock Effects of chiral effective field theory equation of state on binary
  neutron star mergers.
\newblock {\em Phys. Rev. D}, 98:043015, Aug 2018.

\bibitem{Ciolfi17}
R.~Ciolfi, W.~Kastaun, B.~Giacomazzo, A.~Endrizzi, D.~M. Siegel, and R.~Perna.
\newblock {{\it General relativistic magnetohydrodynamic simulations of binary
  neutron star mergers forming a long-lived neutron star}}, 2017.

\bibitem{Ciolfi19}
Riccardo Ciolfi, Wolfgang Kastaun, Jay~Vijay Kalinani, and Bruno Giacomazzo.
\newblock First 100 ms of a long-lived magnetized neutron star formed in a
  binary neutron star merger.
\newblock {\em Phys. Rev. D}, 100:023005, Jul 2019.

\bibitem{KAGRA:2013rdx}
B.~P. Abbott et~al.
\newblock {Prospects for observing and localizing gravitational-wave transients
  with Advanced LIGO, Advanced Virgo and KAGRA}.
\newblock {\em Living Rev. Rel.}, 21(1):3, 2018.

\bibitem{Kastaun16}
B.~Giacomazzo W.~Kastaun, R.~Ciolfi.
\newblock {\it Structure of stable binary neutron star merger remnants: A case
  study}.
\newblock {\em Phys. Rev. D}, 94:044060, Aug 2016.

\bibitem{APR4}
D.~G.~Ravenhall A.~Akmal, V. R.~Pandharipande.
\newblock {\it Equation of state of nucleon matter and neutron star structure}.
\newblock {\em Phys. Rev. C}, 58:1804--1828, Sep 1998.

\bibitem{H4}
N.~K. Glendenning \& S.~A. Moszkowski.
\newblock {\it Reconciliation of neutron-star masses and binding of the
  \ensuremath{\Lambda} in hypernuclei}.
\newblock {\em Phys. Rev. Lett.}, 67:2414--2417, Oct 1991.

\bibitem{LS220}
.~M. Lattimer \& F.~D. Swesty.
\newblock {\it A generalized equation of state for hot, dense matter}.
\newblock {\em Nuclear Physics A}, 535(2):331--376, 1991.

\bibitem{SHT}
S.~Teige G.~Shen, C. J.~Horowitz.
\newblock {\it New equation of state for astrophysical simulations}.
\newblock {\em Phys. Rev. C}, 83:035802, Mar 2011.

\bibitem{BLeos}
Ignazio Bombaci and Domenico Logoteta.
\newblock Equation of state of dense nuclear matter and neutron star structure
  from nuclear chiral interactions.
\newblock {\em Astronomy $\&$ Astrophysics}, 609, 10 2017.

\bibitem{cwb_inj}
{MDC Engine}.
\newblock \url{https://gwburst.gitlab.io/documentation/latest/html/mdc.html}.

\bibitem{Baiotti2008}
L.~Baiotti, B.~Giacomazzo, and Luciano Rezzolla.
\newblock {\it Accurate evolutions of inspiralling neutron-star binaries:
  Prompt and delayed collapse to a black hole}.
\newblock {\em Phys. Rev. D}, 78:084033, Oct 2008.

\bibitem{Goldberg}
J.~N. Goldberg and A.~J. Macfarlane.
\newblock {\it Spin-s Spherical Harmonics and $\partial $}.
\newblock {\em J. Math. Phys.}, 8:2155, Oct 1967.

\bibitem{tesi_Tringali}
Maria~C. Tringali.
\newblock {\em Analysis methods for gravitational wave from binary neutron star
  coalescences: investigation on the post-merger phase}.
\newblock PhD thesis, University of Trento, 2017.

\bibitem{Rezzolla2016}
L.~Rezzolla and K.~Takami.
\newblock {\it Gravitational-wave signal from binary neutron stars: A
  systematic analysis of the spectral properties }.
\newblock {\em Phys. Rev. D}, 93:124051, Jun 2016.

\bibitem{Bauswein2015}
A.~Bauswein and N.~Stergioulas.
\newblock {\it Unified picture of the post-merger dynamics and gravitational
  wave emission in neutron star mergers }.
\newblock {\em Phys. Rev. D}, 91:124056, Jun 2015.

\bibitem{Miani:2023mgl}
Andrea Miani, Claudia Lazzaro, Giovanni~Andrea Prodi, Shubhanshu Tiwari, Marco
  Drago, Edoardo Milotti, and Gabriele Vedovato.
\newblock {Constraints on the amplitude of gravitational wave echoes from black
  hole ring-down using minimal assumptions}.
\newblock 2 2023.

\bibitem{Flaminio}
Raffaele {Flaminio}.
\newblock {Status and plans of the Virgo gravitational wave detector}.
\newblock In {\em Society of Photo-Optical Instrumentation Engineers (SPIE)
  Conference Series}, volume 11445 of {\em Society of Photo-Optical
  Instrumentation Engineers (SPIE) Conference Series}, page 1144511, December
  2020.

\bibitem{AdLIGO2022}
Craig Cahillane and Georgia Mansell.
\newblock {Review of the Advanced LIGO Gravitational Wave Observatories Leading
  to Observing Run Four}.
\newblock {\em Galaxies}, 10(1), 2022.

\bibitem{Hild:2010id}
S.~Hild et~al.
\newblock {Sensitivity Studies for Third-Generation Gravitational Wave
  Observatories}.
\newblock {\em Class. Quant. Grav.}, 28:094013, 2011.

\bibitem{Punturo2010}
{M. Punturo, et al}.
\newblock {The Einstein Telescope: A third-generation gravitational wave
  observatory}.
\newblock {\em Classical and Quantum Gravity}, 27:194002, 2010.

\bibitem{ET2020}
{ET Science Team}.
\newblock Einstein telescope: Science case, design study and feasibility
  report.
\newblock \url{https://apps.et-gw.eu/tds/ql/?c=15662}, 2020.

\end{thebibliography}
\end{document}